\documentclass[twocolumn]{aa}
\bibpunct{(}{)}{;}{a}{}{,}

\usepackage{textcomp}
\usepackage{graphicx}
\usepackage{amssymb}
\usepackage{epstopdf}
\usepackage{gensymb}
\usepackage{natbib}
\usepackage{appendix}
\usepackage{tabu}
\usepackage{color}
\usepackage{multirow}
 \usepackage{ulem}
\usepackage[unicode,colorlinks,citecolor={blue},urlcolor={blue},linkcolor={blue}]{hyperref}
\usepackage[varg]{txfonts}
\usepackage{booktabs}
\usepackage[export]{adjustbox}
\usepackage{threeparttable}
\RequirePackage{color}

\definecolor{darkblue}{rgb}{0.17, 0.49, 0.72}

\definecolor{darkgreen}{rgb}{0.0, 0.4, 0.0}

\newcommand{\kms}{km\,s$^{-1}$}
\newcommand{\he}{\ion{He}{I} 10830\,\AA}
\newcommand{\sref}[1]{Sect.~\ref{#1}}
\newcommand{\fig}[1]{Fig.~\ref{#1}}
\newcommand{\tab}[1]{Table~\ref{#1}}
\newcommand{\app}[1]{Appendix~\ref{#1}}
\makeatletter
\renewcommand*\aa@pageof{, page \thepage{} of \pageref*{LastPage}}
\makeatother

\begin{document}
\title{Magnetized supersonic downflows in the chromosphere}
\subtitle{A statistical study using the \he{} lines}
\institute{Max-Planck-Institut f\"ur Sonnensystemforschung,
Justus-von-Liebig-Weg 3, D-37077 G\"ottingen, Germany\\
\email{krishnamurthy@mps.mpg.de}\label{inst1}
\and
Department of Computer Science, Aalto University, PO Box 15400, FI-00076 Aalto, Finland \label{inst2}
\and
School of Space Research, Kyung Hee University,
Yongin, Gyeonggi 446--701, Korea\label{inst3}
}
\author{K. Sowmya\inst{1} \and A. Lagg\inst{1,2}
\and S. K. Solanki\inst{1,3} \and J.~S. {Castellanos~Dur\'an}\inst{1}}
\date{Received}

\abstract
{The chromosphere above active regions (ARs) on the Sun hosts magnetized supersonic downflows. Studies of these supersonic downflows help to decipher the magnetic fine structure and dynamics of the chromosphere. We perform a statistical analysis of the magnetized supersonic downflows in a number of ARs at different evolutionary stages and survey their characteristics. We analyze spectro-polarimetric scans of parts of 13 ARs obtained in the infrared \he{} triplet formed in the upper chromosphere recorded with the GREGOR Infrared Spectrograph (GRIS) mounted at the GREGOR solar telescope. We retrieve the line-of-sight velocities and the magnetic field vector using the {\sc HeLIx$^+$} inversion code that assumes Milne-Eddington atmospheres. We find magnetized supersonic downflows in all the ARs, with larger area coverage by such flows in ARs observed during their emerging phase. The fact that supersonic downflows were detected in all scans, though they cover only a small fraction, 0.2--6.4\,\%, of the observed field-of-view, suggests that they are a comparatively common phenomenon in the upper chromospheres of ARs. The supersonic downflows are found to be associated with many AR features such as pores, sunspot umbrae, sunspot penumbrae, light bridges, plages, \ion{He}{I} loops as part of arch filament systems characteristic of emerging fields, and filaments. Although several mechanisms are identified to be causing the supersonic downflows, by far the most common one appears to be the draining of plasma along the legs of rising magnetic loops. The loops mainly drain into forming pores. The line-of-sight velocities of the supersonic downflows reach values up to 49\,\kms{} and the velocity distribution shows multiple populations. Almost 92\,\%\ of these supersonic downflows coexist with a subsonic flow component. The weaker, more horizontal fields associated with the supersonic component suggests that it is formed above the subsonic component. 
}

\maketitle

\section{Introduction}
\label{sec:intro}
The chromosphere above active regions (ARs) is known to host fast downflows, which exceed the local sound speed \citep[see e.~g.][]{1995ApJ...441L..51P,2000ApJ...544..567S,2003Natur.425..692S,2004A&A...414.1109L}. Although such supersonic downflows have been seen in the chromosphere, it is still unclear how common they are, nor do we have any robust indication of their average properties. Our focus in this paper is therefore on the statistical exploration of the properties of the velocity and magnetic fields connected with supersonic downflows in the upper chromosphere.

Spectral lines that form at chromospheric heights allow us to probe the plasma dynamics and the magnetic fields in the chromosphere. One such spectral line system is the \he{} triplet formed almost exclusively in a narrow layer in the upper chromosphere \citep{1994IAUS..154...35A,1994A&A...287..229S}. This unusual formation process simplifies the analysis and makes the triplet an extremely valuable tool for studying magnetic and dynamic phenomena in the chromosphere \citep{2017SSRv..210...37L}.

Numerous studies have exploited the diagnostic capabilities of the \he{} lines to probe the chromospheric plasma velocities. \citet{1995ApJ...441L..51P} reported very high velocity redshifts ($30-60$\,\kms{}) in a \he{} filament. Multiple velocity components were seen in a single AR observed with the German Vacuum Tower Telescope by \citet{1997SoPh..172..103M} and in an AR observed with the German Gregory Coud\'e Telescope by \citet{1998ASPC..155..341M}. Using the time sequence of the slit spectra of the \he{} lines that were reported in \citet{1997SoPh..172..103M}, \citet{2000ApJ...544..567S} found signatures of a steady downflow in addition to material almost at rest. 

Chromospheric downflows in emerging flux regions (EFRs), where the sub-surface magnetic flux breaks through the solar surface and forms magnetic loops, have gained increased attention since the study by \citet{2003Natur.425..692S} with an AR observed in \he{} lines. The EFRs appear as dark loops in the chromosphere connecting regions of opposite polarity and are commonly termed arch filament systems (AFS). \citet{2003Natur.425..692S} detected upflows in the loop tops and downflows near the loop footpoints. Using the same observations, \citet{2004A&A...414.1109L,2007A&A...462.1147L} further investigated the velocity structure and obtained downflow speeds of up to 40\,\kms{}. These supersonic downflows in the chromosphere were found to be persistent and to coexist with a second atmospheric component almost at rest within the same spatial resolution element \citep[e.g.,][]{2000ApJ...544..567S,2007A&A...462.1147L,2011ASPC..437..483S, 2018A&A...617A..55G,2019A&A...632A.112Y,2020ApJ...890...82G}.

Supersonic downflows in the \he{} lines (faster than the typical sound speed of 10\,\kms{} at the \he{} triplet formation temperature of about 8000--10000\,K) coexisting with a component nearly at rest always present themselves in the Stokes profiles as an additional redshifted component \citep[e.g., Fig. 9 of ][]{2004A&A...414.1109L}. Attempts were made to fit such intensity profiles to determine the plasma properties using multiGaussian functions \citep{2005ESASP.596E..49A,2007msfa.conf..173A} and multiLorentzian functions \citep{2016AN....337.1057G,2018A&A...617A..55G}. Multicomponent inversions of the full Stokes vector are a more advanced way to analyze the data. They were employed \citep{2007A&A...462.1147L,2010A&A...520A..77X,2011A&A...526A..42S} to analyze both velocity and magnetic fields. We follow this approach and fit all the four Stokes parameters to obtain the magnetic field vector along with the velocity.

Polarization in the \ion{He}{I} lines is generated by the combined action of anisotropic scattering and Zeeman effect \citep{2007ApJ...655..642T}. The effective Land\'e factors of these lines allow for a reliable measurement of magnetic fields with a strength above $50-100$\,G using the Zeeman effect \citep{2017SSRv..210...37L}. Observations of Stokes polarization parameters in the \he{} lines have been used to probe magnetic fields in sunspots \citep[e.g.,][]{1971IAUS...43..279H,1995A&A...293..252R,2006ApJ...640.1153C}, EFRs \citep{2003Natur.425..692S,2004A&A...414.1109L,2007A&A...462.1147L} as well as to investigate weaker magnetic fields in a variety of plasma structures such as chromospheric spicules \citep[e.g.,][]{2005ApJ...619L.191T,2005ApJ...619L.195S,2012ApJ...759...16M}, prominences \citep[e.g.,][]{2006ApJ...642..554M}, and filaments \citep[e.g.][]{1995ApJ...441L..51P,1998ApJ...493..978L,2002Natur.415..403T}.

The analysis of the chromospheric downflows using the \he{} lines have so far been mostly confined to observations of individual ARs (or even just parts of ARs) and a comprehensive study of the characteristics of these downflows is still missing. Here, we analyze the spectropolarimetric scans of parts of 13 ARs and investigate the properties of the magnetized supersonic downflows in the chromosphere. Details of these observations and the data analysis technique are described in \sref{sec:obs}. The detected supersonic downflows are presented in \sref{ar12552}. In \sref{sec:randd} we describe how frequent the magnetized supersonic downflows are, the solar structures they are associated with, along with their velocity and magnetic field distributions. In \sref{sec:sd-disc} we discuss the possible mechanisms through which downflows become supersonic. We present our conclusions in \sref{sec:sum}.
\begin{table*}[htpb]
\centering
\caption{Details of the analyzed raster scan data. See \sref{ssec:datred} and \sref{ssec:sd-vel} for further information.}
\begin{threeparttable}
\begin{tabular}{rrcccccccccc}
\midrule[1.5pt]
\multicolumn{1}{c}{I} & \multicolumn{1}{c}{II} & III & IV & V & VI & VII & VIII & IX & X & XI & XII\\
\midrule
\multicolumn{1}{c}{\multirow{2}{*}{\#}} & \multirow{2}{*}{Dataset ID}& \multirow{2}{*}{NOAA} & Time & $t_{\rm acc}$ & Step & Noise & $x$ & $y$ & \multirow{2}{*}{$\mu$} & AR & Observed part\\
& & & (UT) & (s) & (\arcsec{}) & ($10^{-3}I_c$) & (\arcsec{}) & (\arcsec{}) & & type & of the AR\\
\midrule[1.5pt]
1 & 09may14.008 & 12055 & 11:08 - 11:38 & 2.0 & 0.126 & 1.27 & $-460$ & $+260$ & 0.83 & DAR & Trailing\\
2 & 12may14.006 & 12060 & 09:57 - 10:14 & 3.2 & 0.126 & 0.99 & $-490$ & $-200$ & 0.83 & EAR & Trailing\\
3 & 17jun14.005 & 12087 & 09:13 - 09:27 & 2.4 & 0.126 & 1.25 & $+180$ & $-320$ & 0.92 & EAR & Leading\\
4 & 02sep14.002 & 12152 & 15:15 - 15:32 & 3.2 & 0.124 & 1.02 & $+140$ & $-340$ & 0.92 & EAR & Leading\\
5 & 25may15.002 & 12353 & 08:21 - 08:41 & 1.6 & 0.126 & 1.21 & $+550$ & $+132$ & 0.80 & DAR & Leading\\
6 & 25may15.007 & 12355 & 09:50 - 10:10 & 1.6 & 0.126 & 1.26 & $-660$ & $-143$ & 0.70 & DAR & Leading\\
7 & 01jun15.007 & 12356 & 09:41 - 09:53 & 4.0 & 0.126 & 0.80 & $-250$ & $-252$ & 0.92 & EAR & Leading\\
8 & 09jun16.004 & 12552 & 07:53 - 08:27 & 4.0 & 0.135 & 0.97 & $+573$ & $+254$ & 0.74 & EAR & Leading \& Trailing\\
9 & 19jul16.006 & 12567 & 15:02 - 15:12 & 4.0 & 0.540 & 1.36 & $+250$ & $+000$ & 0.96 & EAR & Trailing\\
10 & 19jul16.008 & 12567 & 15:26 - 15:38 & 4.0 & 0.540 & 1.42 & $+300$ & $+010$ & 0.94 & EAR & Leading \& Trailing\\
11 & 22sep16.005 & 12593 & 10:27 - 11:02 & 4.0 & 0.135 & 0.96 & $+613$ & $+046$ & 0.76 & DAR & Leading\\
12 & 24sep16.000 & 12597 & 09:01 - 09:36 & 4.0 & 0.135 & 0.97 & $+115$ & $-343$ & 0.92 & DAR & Leading\\
13 & 29sep17.011 & 12682 & 14:44 - 15:09 & 4.0 & 0.135 & 1.10 & $-171$ & $-286$ & 0.94 & EAR & Trailing\\
14 & 03oct17.003 & 12683 & 08:32 - 08:58 & 2.4 & 0.130 & 1.45 & $+377$ & $+113$ & 0.91 & EAR & Leading\\
\midrule[1.5pt]
\end{tabular}
\end{threeparttable}
\label{tab:dataset}
\end{table*}

\begin{table*}[ht]
\centering
\caption{Parameters of the spectral lines in the GRIS observing window. }
\begin{tabular}{rrcclcr}
\midrule[1.5pt]
\multirow{2}{*}{Line} & \multirow{2}{*}{Origin} & $\lambda$ & Transition & $g_{\rm eff}$ & $f_{ij}$ & $\Delta\lambda$ \\
& & (\AA{}) & $i-j$ & & & (\kms{})\\
\midrule[1.5pt]
\ion{Si}{I} & Solar & 10827.09 & $4s\,^3{\rm P}_2 - 4p\,^3{\rm P}_2$ & 1.5 & 3.47e-01 & $-89.96$\\
\ion{He}{I}a & Solar & 10829.09 & $2s\,^3{\rm S}_1 - 2p\,^3{\rm P}_0$ & 2.0 & 5.99e-02 & $-34.60$\\
\ion{Ca}{I} & Solar & 10829.27 & $4p\,^3{\rm F}_3 - 6d\,^3{\rm D}_2$ & 1.0 & 0.42e-02 & $-29.62$\\
\ion{He}{I}b & Solar & 10830.25 & $2s\,^3{\rm S}_1 - 2p\,^3{\rm P}_1$ & 1.75 & 1.78e-01 & $-2.49$\\
\ion{He}{I}c & Solar & 10830.34 & $2s\,^3{\rm S}_1 - 2p\,^3{\rm P}_2$ & 1.25 & 2.99e-01 & 0.00\\
H$_2$O & Telluric & 10831.63 & - & - & - & 35.70\\
H$_2$O & Telluric & 10832.09 & - & - & - & 48.44\\
\ion{Ca}{I} & Solar & 10833.38 & $4p\,^3{\rm P}_1 - 3d\,^3{\rm P}_2$ & 1.5 & 1.02e-01 & 84.14\\
H$_2$O & Telluric & 10834.00 & - & - & - & 101.31\\
\ion{Na}{I} & Solar & 10834.84 & $3d\,^2{\rm D}_{5/2} - 6f\,^2{\rm F}_{5/2}$ & 1.03 & 2.62e-03 & 124.56\\
\ion{Na}{I} & Solar & 10834.84 & $3d\,^2{\rm D}_{5/2} - 6f\,^2{\rm F}_{7/2}$ & 1.07 & 5.24e-02 & 124.56\\
\ion{Na}{I} & Solar & 10834.90 & $3d\,^2{\rm D}_{3/2} - 6f\,^2{\rm F}_{5/2}$ & 0.90 & 5.50e-02 & 126.22\\
H$_2$O & Telluric & 10837.97 & - & - & - & 211.20\\
\ion{Ca}{I} & Solar & 10838.97 & $4p\,^3{\rm P}_2 - 3d\,^3{\rm P}_2$ & 1.5 & 1.85e-01 & 238.88\\
\midrule[1.5pt]
\end{tabular}
\label{tab:lambda}
\end{table*}
\section{Observations and data analysis}
\label{sec:obs}
\subsection{Data and reduction}
\label{ssec:datred}
The data were recorded with the GREGOR Infrared Spectrograph \citep[GRIS;][]{2012AN....333..872C} mounted at the 1.5\,m GREGOR solar telescope \citep{2012AN....333..796S,2012AN....333..810D} at the Observatorio del Teide. The observations, summarized in \tab{tab:dataset}, consist of 14 spectropolarimetric raster scans of parts of 13 ARs, recorded between 09 May 2014 and 03 October 2017. Columns denote the number we give to each dataset (i.e., scans of the slit across the solar surface; column I), the ID of the scan as used in the GRIS archive\protect\footnote{\url{http://archive.leibniz-kis.de/pub/gris/}} (column II), NOAA number of the target regions (column III), the start - end time of the scan (column IV), the accumulation time per slit position to record one full Stokes vector (column V), the step size of the scan (column VI), the average noise for the data binned over two spectral pixels (average of the noise in $Q, U, V$; in units of $I_c$, which is the continuum intensity; column VII), the heliocentric $x$ (column VIII) and $y$ (column IX) coordinates of the center of the scanned field-of-view (FOV), the cosine of the heliocentric angle $\theta$ ($\mu={\rm cos}\,\theta$) of the AR (column X), the evolutionary stage of the AR (EAR - emerging; DAR - decaying) at the time of the scan (column XI) and whether the leading or the trailing part of the AR was observed (column XII). A description of the observed ARs and their intensity images are given in Appendix~\ref{sec:app-desc}. The pixel size along the slit is 0.135\arcsec{}, and the step size chosen for the scanning was set close to an integer multiple of this value ($\times 1$ or $\times4$). The number of accumulations and the exposure time per slit position were in the range $4-20$ and $30-100$\,ms, respectively.

\begin{figure*}[htpb]
    \centering
    \includegraphics[scale=0.9,trim=0.0cm 0.cm 0.0cm 5.0cm,clip]{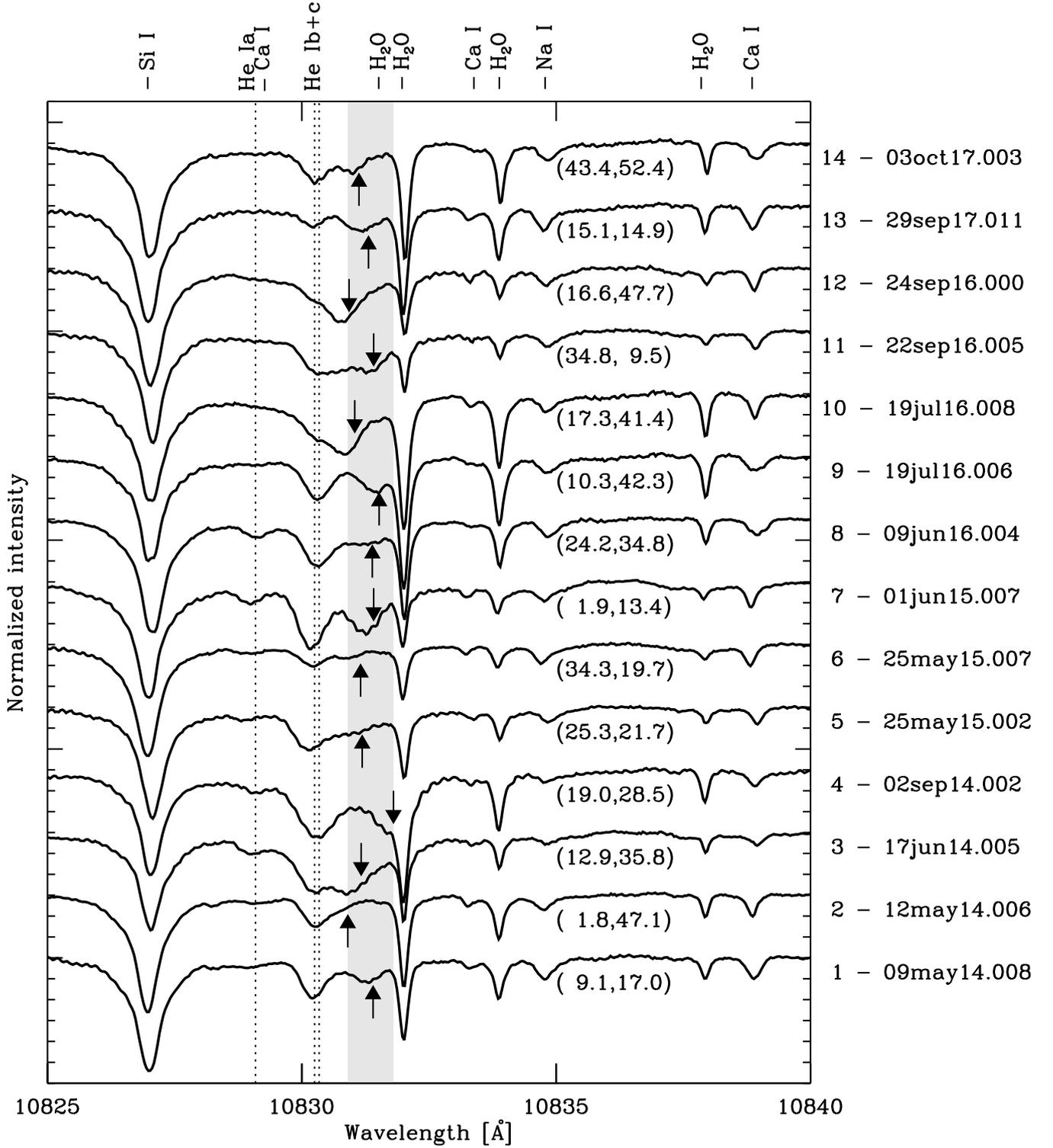}
    \caption{Examples of Stokes $I/I_c$ profiles, one from each of the 14 datasets, indicating the presence of a strongly redshifted component of the \he{} triplet, marked by the arrows. The arrows shown here span the wavelength range corresponding to a velocity shift between 15.6 and 40.4\,\kms{} from the \ion{He}{I}c line center. This wavelength range is indicated by the light gray shaded area. The vertical black dotted lines mark the rest wavelengths of the three transitions of the \he{} triplet (see \tab{tab:lambda}). The numbers in brackets are the $(x,y)$ coordinates in arcseconds of location of these profiles on the corresponding intensity images given in \app{sec:app-desc}. The locations of the profiles shown here are also indicated with black crosses on the corresponding He intensity images (panels b) in  Figs.~\ref{fig:app-maps1} -- ~\ref{fig:app-maps4}. The other spectral lines in the observed window are identified at the top of the frame. The labels on the right give the dataset number (see column I of \tab{tab:dataset}) and the dataset ID (see column II of \tab{tab:dataset}) to which the profiles belong.}
    \label{fig:down}
\end{figure*}

The observed spectral region covers a range from 10823 to 10842\,\AA{} with a sampling of about 18\,m\AA{}/pixel. The \he{} triplet consists of three spectral lines: \ion{He}{I}a at 10829.0911\,\AA{}, \ion{He}{I}b at 10830.2501\,\AA{}, and \ion{He}{I}c at 10830.3397\,\AA{}. \ion{He}{I}b and \ion{He}{I}c are blended as their wavelength separation is smaller than their widths. In addition to the chromospheric \ion{He}{I} triplet, the two photospheric lines, \ion{Si}{I} 10827\,\AA{} and \ion{Ca}{I} 10839\,\AA{}, and a telluric blend at 10832\,\AA{} are prominent examples of the other lines present in the observed spectral region.  \tab{tab:lambda} provides the parameters of all the spectral lines identified on \fig{fig:down}. The table distinguishes between the spectral lines of solar origin and the telluric lines (which arise due to absorption in the Earth's atmosphere). The lower ($i$) and the upper ($j$) atomic terms involved in the transition, the effective Land\'e factor ($g_{\rm eff}$), the oscillator strength ($f_{ij}$) and the wavelength separation with respect to the \ion{He}{I}c line ($\Delta\lambda$) in velocity units are also provided in \tab{tab:lambda}.

The dark current subtraction, flat-fielding, polarimetric calibration and cross-talk removal were done with the standard GRIS data reduction software. The image resolution, calculated by averaging the power spectrum along the slit direction, lies in the range 0.4\arcsec{}-0.8\arcsec{}. The diffraction limited resolution of GREGOR at 10830\,\AA{} is 0.18\arcsec{}. However, due to the seeing conditions, and non AO-corrected aberrations over the time of accumulating one full Stokes vector (between 1.6 and 4 seconds) this resolution could not be achieved. A spectral atlas obtained by the Fourier Transform Spectrometer \citep[FTS;][]{1984SoPh...90..205N} at the McMath-Pierce solar telescope on Kitt Peak was used for the wavelength calibration and continuum correction. The continuum correction curve in the form of a polynomial, was obtained by fitting the FTS spectrum to the profile computed from the GRIS flat field images, excluding the telluric and the solar lines. Wavelength offset, dispersion, spectral resolution and spectral straylight are among the free parameters of the fitting procedure. We remark that the reduction procedure still leaves some residual fringes in Stokes $I$ as can be seen in \fig{fig:down}. These fringes, however, do not affect our analysis as they are far narrower than the widths of solar spectral lines.

Intensity maps of each dataset at \ion{He}{I}c line center and at a continuum wavelength blueward of the \ion{Si}{I} line are presented in \app{sec:app-desc}. Note that some maps were recorded without the image derotator installed at GREGOR and are therefore slightly distorted. This is due to the image rotation during the scanning period caused by the altitude-azimuth mount of the telescope. In order to increase the S/N, all spectra were binned over two spectral pixels. The noise for the Stokes polarization parameters was computed as follows: we chose a wavelength region in the continuum (blueward of the \ion{Si}{I} 10827\,\AA{} line) and calculated the standard deviation of the Stokes signals over wavelength for each spatial pixel. The mean values of the standard deviations over spatial pixels were calculated for each of the Stokes $Q$, $U$, and $V$ parameters and this mean was taken as the 1$\sigma$ level. The noise levels computed in this way for each dataset analyzed in this paper are given in \tab{tab:dataset}. For the analysis in this paper we have, however, used the noise in the individual Stokes parameters.

Using the continuum images and line-of-sight (LOS) magnetograms from the Helioseismic and Magnetic Imager \citep[HMI;][]{2012SoPh..275..207S} onboard the Solar Dynamics Observatory \citep[SDO;][]{2012SoPh..275....3P}, we tracked the evolution of the analyzed ARs over three consecutive days (the day of observation $\pm1$\,day). Based on the evolution of the size of the AR in the continuum images and appearance/disappearance of magnetic elements we classified the observed ARs into emerging (EAR) and decaying (DAR; see column XI of \tab{tab:dataset}).

\subsection{Inversions}
\label{ssec:inversion}
Thanks to the formation of the \he{} lines exclusively in the upper chromosphere, new insights into the velocity and magnetic structure of the upper chromosphere have been obtained in recent years using ground-based observations. Being formed in a highly-corrugated layer, on average at about 2000 -- 2400\,km above the visible solar surface \citep{1994IAUS..154...35A,1994A&A...287..229S}, the profiles of the \he{} triplet have almost no contamination from photospheric light. This is due to the particular formation mechanism of these lines. \he{} infrared lines originate from transitions involving the triplet state of He (orthohelium). The majority of the He population lies in the singlet state (parahelium) under normal chromospheric conditions leaving the triplet state electron deprived. Photoionization from coronal extreme ultra-violet (EUV) radiation from above and collisional excitation ionizes the He atoms. The ionized He atoms then recombine and de-excite with equal probability to either the singlet or triplet state \citep{1994IAUS..154...35A,2017SSRv..210...37L}. Due to the large density difference between the chromosphere and the corona, the chromosphere is highly opaque to the coronal EUV radiation. This restricts the photoionization and hence the \he{} triplet transitions to a thin layer at the top of the chromosphere \citep[e.g.,][]{2016A&A...585A...4C}. Therefore, the physical parameters within this layer can be assumed to be height independent, which makes it possible to carry out radiative transfer modeling in Milne-Eddington-type atmospheres.

To analyze the spectropolarimetric data chosen for our study, we used the {\sc HeLIx$^+$} inversion code \citep{2004A&A...414.1109L,2009ASPC..415..327L}, based on the Unno-Rachkowsky solution of the radiative transfer equation. Line inversions were carried out in a Milne-Eddington-type atmosphere taking into account the Zeeman effect and the incomplete Paschen--Back effect \citep{2005ApJS..160..312S,2006A&A...456..367S}. Such inversions allow us to retrieve the full magnetic vector and the LOS velocity in the upper chromosphere where the \he{} triplet is formed. The \ion{Si}{I} 10827\,\AA{} line (formation height of the line core $\sim 500$\,km above the visible solar surface; \citealt{2017A&A...603A..98S}) and the \ion{Ca}{I} 10839\,\AA{} line (formation height $\sim 150$\,km above the visible solar surface) are also simultaneously inverted with {\sc HeLIx$^+$} to obtain the magnetic field vector and LOS velocity in the upper and lower photosphere, respectively. The \ion{He}{I}a line overlaps with the broad wing of the \ion{Si}{I} line as seen in \fig{fig:down}. It is therefore necessary to invert the \ion{Si}{I} and \ion{He}{I} lines simultaneously in order to obtain reliable fit to the He lines. In a visual analysis of our sample, we did not find any supersonic downflows located far beyond the telluric line at 10832\,\AA{}. Therefore we excluded the wavelength region 10833 - 10838\,\AA{} from the fit. 

The free parameters of the inversion are: the magnetic field strength, $B$, the inclination of the magnetic field to the LOS, $\gamma$, and its azimuth, $\phi$, the LOS velocity, $v_{\rm LOS}$ (used interchangeably with $v$ throughout the paper), the damping constant, $a$, the ratio of line-center-to-continuum opacity, $\eta_{\rm 0}$, and the gradient of the source function, $S_{\rm 1}$. A set of these 8 parameters constitute an atmospheric component. A separate set of such parameters is obtained for each layer of the atmosphere, i.~e. for the upper chromosphere (from the \ion{He}{I} triplet), the upper photosphere (from the \ion{Si}{I} line) and the lower photosphere (from the \ion{ca}{I} line).

Fast downflows are often associated with a separate component of the atmosphere that is nearly at rest within the same spatial resolution element. Some examples of profiles where we see two components in \ion{He}{I} are shown in \fig{fig:down}. To properly account for the two coexisting atmospheric components, we inverted all observed maps using two magnetized atmospheric components (slow and fast) for the \he{} triplet. The slow and the fast components have the same range of values for the free parameters except for the LOS velocity. The filling factor, $\alpha$, denotes the contribution from the fast component to the total observed profile, while $1-\alpha$ is the contribution from the slow component. The criteria we employ to distinguish the two components are outlined in \sref{ssec:criteria}. In addition, we also inverted all data using a single atmospheric component. We used this to test how well the inversions using two atmospheric components agreed with the one-component inversions for regions where no second component is clearly visible or where the two components are not well separated. We find that in such cases the results for the component with a larger filling factor in the two component inversions agree well with those of the single component inversions. Therefore, we use the parameters from the two component inversions for all regions analyzed (see \sref{ssec:criteria} for details).

The three \ion{He}{i} lines, formed under identical conditions, require only one identical set of atmospheric parameters per component \citep[see][for further details]{2004A&A...414.1109L}. Beyond that we fixed the Doppler width for the \ion{He}{I} lines in order to reduce the number of free parameters. This increased the stability of the minimization process without decreasing the quality of the fit. In addition to the two magnetized atmospheres for the \ion{He}{I} lines, a magnetized and a nonmagnetized component for the \ion{Si}{I} line and a magnetized component for the \ion{Ca}{I} line were used. The nonmagnetized component for the \ion{Si}{I} line is needed to obtain satisfactory fits to the broad wings of the \ion{Si}{I} line without contaminating the magnetic field information obtained from the line core. The telluric blend at 10832\,\AA{} was also fitted using a Voigt profile.

We use the LOS velocity maps for the magnetized \ion{Si}{I} component to calibrate the velocities of the \ion{He}{i} lines (see \sref{ssec:criteria}). The calibration is performed by choosing a quiet Sun region within each dataset and setting the mean velocity in this region computed from the magnetic component of the \ion{Si}{I} line to zero. We define the slow component $v_{\rm LOS}$ to lie within $\pm10$\,\kms{} around this zero velocity. This should ensure the slow component to be subsonic and also exclude supersonic upflows, which are not the subject of this paper. The fast component is defined to be strictly supersonic and takes $v_{\rm LOS}$ values between 10 and $50$\,\kms{}. A visual inspection of the observed Stokes profiles indicated that there are no downflows faster than 50\,\kms{} in our sample and hence this value is chosen as the upper limit for the fast component velocity. 

\begin{figure}[ht!]
    \centering
    \includegraphics[scale=0.75,trim=3.2cm 0.cm 5.0cm 12.0cm,clip]{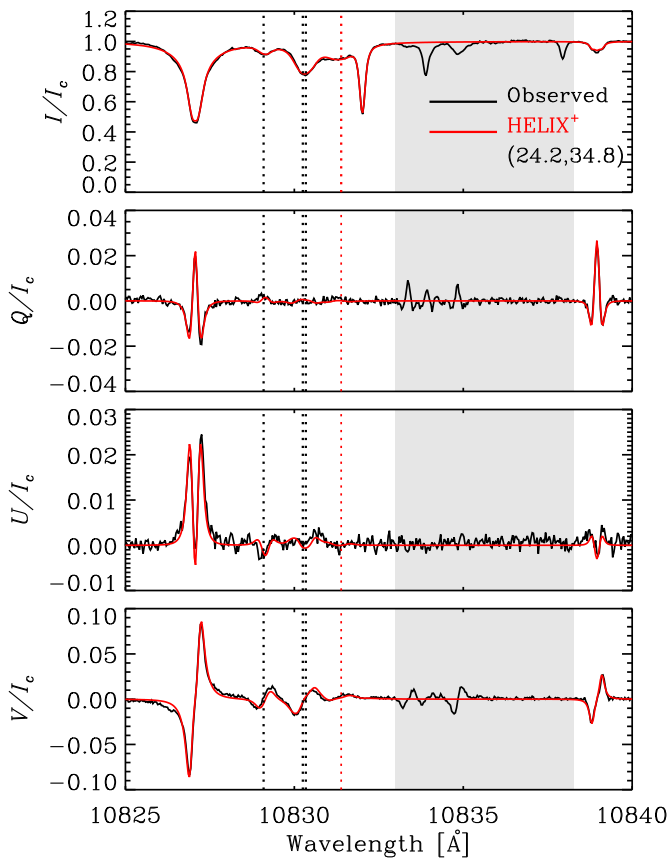}
    \caption{A comparison between the observed Stokes profiles (black) and the best fit found by the {\sc HeLIx$^+$} code (red) using two magnetized atmospheric components for the \he{} triplet, a magnetized and a straylight component for the \ion{Si}{I} 10827\,\AA{} line and a magnetized component for the \ion{Ca}{I} 10839\,\AA{} line. The telluric line at 10832\,\AA{} is fitted using a Voigt function. The grey shaded area marks the wavelength region that is not fitted. The vertical red dotted lines indicate the central position of the redshifted flow component with $v_{\rm LOS}$ of about 29\,\kms. The vertical black dotted lines and numbers in parenthesis in the Stokes $I/I_c$ panel have the same meaning as in \fig{fig:down}. The observed Stokes $I/I_{c}$ profile shown here is the same as the profile labeled 8 -- 09jun16.004 in \fig{fig:down}.}
    \label{fig:stkprf}
\end{figure}

The capabilities of the {\sc HeLIx$^+$} code in terms of how well it can fit the observed Stokes profiles have been demonstrated in earlier papers \citep[e.g.][]{2004A&A...414.1109L,2007A&A...462.1147L,2011A&A...526A..42S}. An example of the Stokes profiles synthesized by using two atmospheric components for the \he{} lines and how they compare with the observed Stokes profiles is given in \fig{fig:stkprf} for the intensity profile labeled 8 - 09jun16.004 in \fig{fig:down}.

\begin{figure*}[htpb]
    \centering
    \includegraphics[scale=0.8,trim=0.0cm 4.0cm 0.cm 4.cm,clip]{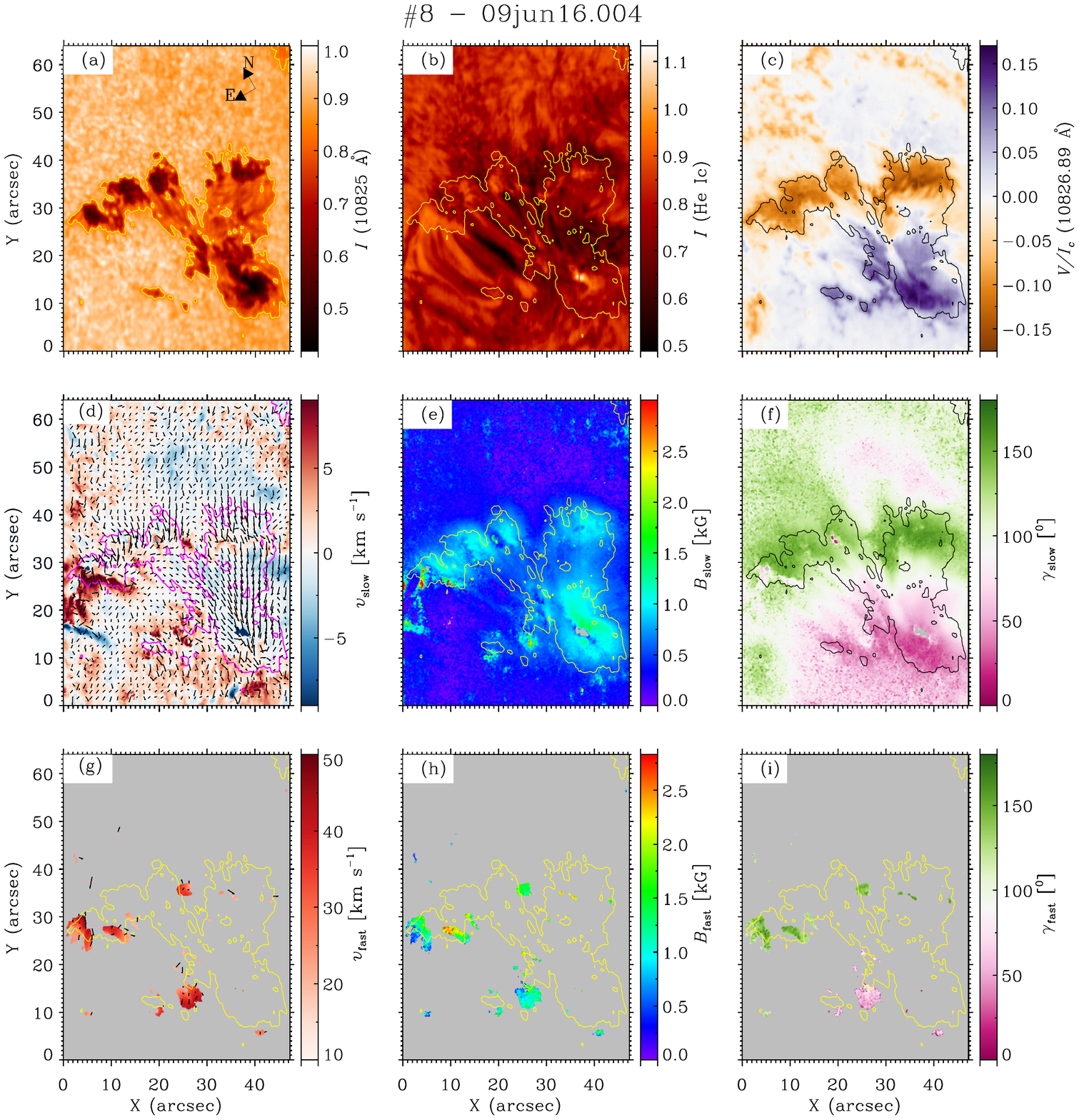}
    \caption{Results of the two component inversions of the \ion{He}{I} lines for AR\,12552 observed on 09 June 2016 (dataset 8). Panels (a)--(c) represent the normalized intensities at 10825\,\AA{} and at \ion{He}{I}c line center wavelengths, and Stokes $V/I_c$ map at 10826.89\,\AA{}, respectively. The LOS velocity, the strength and the inclination of the magnetic field are presented in panels (d)--(f) for the slow and in panels (g)--(i) for the supersonic components. The contours mark the boundaries of sunspots and pores, determined using the continuum intensity image shown in panel (a). These contours enclose regions within 85\,\% level of the surrounding quiet Sun intensity in the continuum image. The arrows in panel (a) indicate solar north and east directions. The azimuths of the magnetic field are over-plotted as lines in panels (d) and (g). See \sref{ar12552} for further details.}
    \label{fig:invmaps}
\end{figure*}
\subsection{Classification of the observed regions}
\label{ssec:class-obs}
At this point, we remark that two telluric lines are present at wavelengths corresponding to velocity shifts of $\sim$36 and 48\,\kms{} from the \ion{He}{I}c line (see \tab{tab:lambda}). Even though the {\sc HeLIx$^+$} code takes the telluric lines into account, it is still possible that a second component is wrongly placed at the location of the telluric line. For this reason, we do not trust the results of the inversion when the supersonic downflow signature is seen only in Stokes $I$ and hence restrict our analysis to magnetized supersonic downflows, where the polarized profiles unambiguously identify the downflows to be solar.

The magnetic field expands rapidly with height, forming a magnetic canopy in the lower to middle chromosphere \citep{1982SoPh...79..267G,1982SoPh...79..247J,1990A&A...234..519S}, so that by the height of formation of the \ion{He}{I} triplet, the magnetic field fills the whole atmosphere and hence all flows are expected to be magnetized to some degree. A preliminary investigation of the observed Stokes profiles indicates that the magnetized supersonic downflows, which are the subject of this paper, cover only a small fraction of the observed FOV. In order to quantify the percentage of pixels that show magnetized fast downflows in a given dataset, we need a reference parameter that is independent of the size of the scanned region. The total number of pixels in the observed FOV cannot be used as a reference as some datasets have dominantly AR (or a part of the AR) in the FOV while a significant part of the FOV of other scans is covered by the quiet Sun where the magnetic field is weak and often is hard to detect in many of the pixels. Therefore, we decided to classify the pixels in the FOV for each dataset into magnetized and field-free. The latter are all pixels with a polarization level lower than 3.5$\sigma$ in the observed Stokes $Q$, $U$, and $V$ in the slow component of \ion{He}{I}. A similar approach was used by \citet{2007msfa.conf..173A}. This categorization will enable us to quantify how many of the pixels magnetized in the slow He component also display a magnetized supersonic downflow component.

For this purpose, we first determine the wavelength position of the slow He component at each spatial pixel using the LOS velocity map from the inversions. Then we check if the parameter $M$, defined as $M = (\max{|Q|}~or~\max{|U|}~or~\max{|V|}$), is greater than 3.5$\sigma$ for the corresponding Stokes parameter within $\pm0.36$\,\AA{} from the central position of the slow component. When the filling factor of the slow component is less than 20\,\%, we check if the magnetic signal of the fast component is above the threshold imposed. We then classify all those pixels having $M>3.5\sigma$ as magnetized and those with $M\le3.5\sigma$ as field-free. We carry out this classification initially using the observed profiles ($M_O$). The total number of pixels (profiles) in the FOV (N$_{\rm prof}$) and the percentage of pixels classified to be magnetized in the observed data are listed in columns II and III of \tab{tab:pixel-wlbin}.

In the observed Stokes profiles, the magnetic signals of the slow and the fast components are often superimposed, especially when the velocity separation between the two components is small. This makes it challenging to distinctly identify the magnetic signals of the two components. In the inversions, however, the synthetic profiles of the slow and the fast components are computed separately and then combined to match the observed profiles. It is therefore advantageous to use the synthetic profiles of the individual components to determine if their polarized signals are sufficiently above the noise or not. Making use of the synthetic profiles and the $\sigma$ of the observed Stokes parameters, we calculate the percentage of pixels with the parameter $M_S>2\sigma$ in a given dataset. These numbers are given in column IV of \tab{tab:pixel-wlbin}. We find that the 3.5$\sigma$ criterion for the observed profiles to be classified as magnetized is similar to the 2$\sigma$ criterion for synthetic profiles (compare columns III and IV in \tab{tab:pixel-wlbin}). It is clear from \tab{tab:pixel-wlbin} that $M_S>2\sigma$ gives a tighter criterion (fewer pixels) than $M_O>3.5\sigma$. Observed profiles give a larger percentage of magnetized pixels when we impose the 2$\sigma$ criterion on them. This stringent 2$\sigma$ threshold cannot be applied to the observed profiles, since under the assumption of a Gaussian noise characteristics over the 63 measurement points (21 points each for the Stokes $Q$, $U$, and $V$), it would mean that virtually all pixels get classified as magnetized. With the advantage of being able to separate the signals of the slow and the fast components, the 3.5$\sigma$ criterion on the observed profiles can be replaced by the 2$\sigma$ criterion on the synthetic profiles. According to this scheme, all pixels with $M_S>2\sigma$ are magnetized. The pixels with $M_S\le2\sigma$ are not considered in our analysis.

Note that in \tab{tab:pixel-wlbin} the percentage of profiles with $M_S>2\sigma$ in dataset 8 is much lower than those with $M_O>3.5\sigma$. This difference is perhaps due to the pixels which show Stokes $Q$ and $U$ signals, with either of them being just strong enough to satisfy the $M_O>3.5\sigma$ criterion (but with Stokes $V$ close to the noise level). Since the fit to such profiles is poor, the Stokes $Q$ and $U$ signals in the synthetic profiles do not meet the $M_S>2\sigma$ threshold. For this dataset, the $M_S>2\sigma$ criterion turns out to be similar to $M_O>4\sigma$.

\subsection{Criteria for identification of magnetized fast downflows}
\label{ssec:criteria}
In some of the regions classified as field-free (i.e. regions with $M_S\le2\sigma$ as described in \sref{ssec:class-obs}), supersonic velocities were retrieved in the \ion{He}{I} lines. They mostly correspond to regions where the inversion code wrongly places a second He component at the location of the telluric line at 48\,\kms{}. Some regions in the observed maps show fast magnetized downflows alone while some others show their association with an additional component almost at rest. In regions where we find such dual flows, we applied the following criteria to the synthetic profiles in order to identify the magnetized supersonic downflows:
\begin{itemize}
    \item[a)] the two components must be separated sufficiently well from each other. Therefore, we require the fast component to be shifted by more than 10\,\kms{} relative to the slow component and the fast component velocity should be greater than 10\,\kms{} (supersonic);
    \item[b)] the filling factors of each of the two components must be at least 20\,\%\ for this to be classified as a dual flow;
    \item[c)] the peak value of any of the three synthetic Stokes parameters $|Q|$~or~$|U|$~or~$|V|$ in a range of $\pm0.36$\,\AA{} centered at the fast component position must be larger than a given threshold in their respective values (i.e., $2\times\sigma(Q)$~or~$2\times\sigma(U)$~or~$2\times\sigma(V)$).
\end{itemize}

In observations where fast magnetized downflows existed with another component having a filling factor less than 20\,\%\, we checked only for the last criterion listed above. The requirement imposed on the magnetic signal (criterion c) ensures that unmagnetized supersonic downflows (which may be affected by the telluric lines as discussed above) are excluded from our analysis.

The number of profiles in each dataset identified to have magnetized supersonic downflows (N$_{\rm sd}$), using the criteria outlined above, is given in column V of \tab{tab:pixel-wlbin}. Column VI gives the fraction of pixels with clear chromospheric magnetic signals also display a supersonic downflow for each dataset. The maximum velocities of the magnetized supersonic downflows are given in column VII. Column VIII represents the mean and the standard deviation of these supersonic velocities.

\begin{table*}[ht]
\centering
\caption{Statistics from the classification scheme described in \sref{sec:obs}. Datasets marked with "$\star$" are classified as DARs. See \sref{sec:randd} for further discussion.}
\begin{tabular}{rrccrrcc}
\midrule[1.5pt]
\multicolumn{1}{c}{I} & \multicolumn{1}{c}{II} & III & IV & \multicolumn{1}{c}{V} & \multicolumn{1}{c}{VI} & VII & VIII \\
\midrule
\multicolumn{1}{c}{\multirow{2}{*}{\#}}  & \multicolumn{1}{c}{\multirow{2}{*}{N$_{\rm prof}$}} & \% of profiles & \% of profiles & \multicolumn{1}{c}{\multirow{2}{*}{N$_{\rm sd}$}}
& \multicolumn{1}{c}{\multirow{2}{*}{\%}} & max($v_{\rm fast}$) & $\langle v_{\rm fast}\rangle\pm\sigma_{v_{\rm fast}}$ \\
& & with $M_O>3.5\sigma$ & with $M_S>2\sigma$ & & & (\kms{}) & (\kms{}) \\
\midrule[1.5pt]
$^\star${1} & { 141000} &  { 72.00} &  { 71.03} & { 455} & { 0.45} & { 33.2} & {$21.5\pm5.4$} \\
2  & 46900 &  98.40 &  96.99 & 2918 & 6.41 & 46.4 & $23.2\pm7.1$ \\
3  & 69881 &  57.86 &  55.44 & 355 & 0.92 & 36.4 & $25.2\pm4.9$ \\
4  & 94000 &  81.28 &  76.66 & 2688 & 3.73 & 49.2 & $27.2\pm7.7$ \\
$^\star${5}  & { 201600} &  { 33.70} &  { 28.12} & { 168} & { 0.29} & { 40.2} & {$22.8\pm5.4$} \\
$^\star${6}  & { 203850} &  { 15.38} &  { 13.08} & { 94} & { 0.36} & { 29.8} & {$22.5\pm4.2$} \\
7  & 45200 &  74.41 &  73.88 & 157 & 0.47 & 47.2 & $25.0\pm4.8$ \\
8  & 165900 &  72.14 &  58.00 & 4479 & 4.65 & 43.4 & $22.7\pm7.2$ \\
9  & 49400 &  57.24 & 56.24 & 394 & 1.42 & 37.9 & $23.9\pm7.5$ \\
10  & 57000 &  67.33 &  67.11 & 157 & 0.41 & 42.5 & $26.3\pm9.4$ \\
$^\star${11}  & { 169200} &  { 47.92} &  { 42.91} & { 267} & { 0.36} & { 33.4} & {$23.0\pm5.6$} \\
$^\star${12}  & { 171000} &  { 40.74} &  { 36.22} & { 116} & { 0.18} & { 22.8} & {$15.2\pm2.5$} \\
13  & 109000 &  69.96 &  69.95 & 472 & 0.62 & 39.2 & $25.2\pm8.0$ \\
14  & 170800 &  57.03 &  53.42 & 746 & 0.82 & 37.3 & $21.4\pm7.6$ \\
\bottomrule[1.5pt]
\end{tabular}
\label{tab:pixel-wlbin}
\end{table*}
\section{Description of magnetized supersonic downflows in a selected dataset}
\label{ar12552}
In this section, we present in greater detail the results for one dataset that harbors nearly the entire range of profiles and solar features found to show fast magnetized downflows, namely dataset 8. A detailed analysis of the downflows in dataset 3 from 17 June 2014, which shows a \ion{He}{I} filament associated with AR\,12087 is presented in \citet{2020IAUS..354..454S}. The statistical properties of such downflows in all the regions analyzed are discussed in \sref{sec:randd}.

The dataset 8 was observed on 09 June 2016. It consists of the scan of AR\,12552 and the scan lasted for about 35 minutes starting at 07:53 UT, while the AR was still emerging on the northwestern side of the visible hemisphere. The scan covers about 47\arcsec{} in the scan direction ($x$-axis) and 64\arcsec{} along the slit ($y$-axis), with a step size of 0.135\arcsec{} and a pixel size of 0.135\arcsec{} along the slit. The center of the FOV corresponds to a heliocentric angle of $42\degree$ ($\mu=0.74$, $x=+573$\arcsec{} and $y=+254$\arcsec{}, see \tab{tab:dataset}). Panels (a)--(c) in \fig{fig:invmaps} show intensity maps at 10825\,\AA{} in the local continuum, at 10830.3\,\AA{} (\ion{He}{I}c), and Stokes $V/I_c$ at 10826.9\,\AA{}, respectively. A sunspot group is visible in the continuum intensity image, along with pores. The following positive polarity (purple patch in \fig{fig:invmaps}c) has formed into a single sunspot whereas the leading negative polarity (brown patch in \fig{fig:invmaps}c) has fragmented into four spots whose penumbrae are still forming. An arcade of loops connecting opposite polarities (forming a so-called AFS) is visible in the \ion{He}{I}c intensity image. A small patch where \ion{He}{I} goes into emission is seen on the lower right of \fig{fig:invmaps}b. We exclude these pixels from our analysis as the inversion was not optimized to fit emission profiles.

The results of the two-component inversions are shown in panels (d)--(i) in \fig{fig:invmaps}. The LOS velocity, the magnetic field strength, and the inclination of the magnetic field for the slow component are shown in panels (d)--(f). Panels (g)--(i) show, respectively, the LOS velocity, the magnetic field strength and the inclination of the magnetic field for the fast component at only those locations where the presence of supersonic downflows was inferred using the conservative criteria outlined in \sref{ssec:criteria}. The contours mark the boundaries of sunspots and pores seen in the continuum intensity map in panel (a). These photospheric contours are obtained by setting a threshold of 85\,\%\ of the surrounding quiet Sun intensity in the continuum intensity map (see also \app{sec:app-desc}). The magnetic field azimuths are over-plotted as black lines in panels (d) and (g). All maps have been smoothed with a median filter, for representation purposes. 

The loop tops show subsonic blue shifts in the LOS velocity as seen in panel (d), indicating that the loops are rising and transporting relatively cool material from the surface to the chromosphere. Magnetized supersonic downflows are found to exist at the footpoints of the loops (see panel g). These downflows attain a maximum velocity of 43.4\,\kms{} and are located close to pores, nearby plage regions, sunspot umbrae and penumbrae. The average supersonic downflow velocity is 22.7\,\kms{} with a standard deviation of 7.2\,\kms{}. The magnetized supersonic downflows fill only a small fraction of the observed FOV (compare panels d and g). Out of the 58\,\%\ of pixels in the FOV which are classified as magnetized, only 4.65\,\%\ show the presence of a supersonic downflow component (see columns V and VI of \tab{tab:pixel-wlbin}). The magnetic vector of the slow and the fast components are slightly different in regions where the fast component is found. The chromospheric magnetic field strengths corresponding to the slow and fast components peak at 641 and 458\,G, respectively (with the corresponding median values of 629 and 527\,G). The chromospheric LOS inclinations peak at 59 and 68\degree\ for the slow and the fast components (the corresponding median values are 54 and 55\degree).

\section{Properties of magnetized supersonic downflows}
\label{sec:randd}
In our analysis with two component inversions, we identified both single component magnetized supersonic downflows (with the slow component having a filling factor below 20\,\%, which typically means that it is hardly visible above the noise) and supersonic downflows which coexist with a subsonic component (which can be an upflow or a downflow). In the following, we discuss the properties of such flows.

\subsection{Frequency of occurrence}
\label{ssec:sd-perc}
As discussed in \sref{ssec:criteria}, the number of profiles exhibiting magnetized supersonic downflows and their fraction with respect to the total number of magnetized pixels in the FOV are given in \tab{tab:pixel-wlbin} (columns V and VI). Recall that to determine whether a pixel is magnetized we require $M_S>2\sigma$. There are, in total, 13466 profiles from the 14 scans displaying magnetized supersonic downflows. Of these, 1067 ($\sim8$\,\%) are single component profiles i.e. the overwhelming majority of supersonic downflows are associated with a component close to rest. Although the area coverages are small ($0.2-6.4$\,\%), we find supersonic downflows associated with magnetic fields in all ARs studied, indicating that they are a common phenomenon in ARs. 

It is apparent from \tab{tab:pixel-wlbin} that the relative area covered by magnetized supersonic downflows is higher in EARs than in DARs (marked with $\star$ in column I of \tab{tab:pixel-wlbin}). Nine EARs in our sample account for 91.8\,\%\ (N$_{\rm sd}$=12366) of the total profiles with magnetized supersonic downflows, while five DARs together contribute to only 8.2\,\%\ (N$_{\rm sd}$=1100) of such profiles (we note that this lower fraction could also partly be due to the fewer number of scans of DARs). Overall, the area covered by supersonic downflow-hosting pixels is only 1.78\,\%\ and 0.19\,\%\ of the area of the EARs and DARs, respectively. This confirms that supersonic downflows are indeed much more common in EARs than in DARs.

One of the 13 ARs in our sample, namely AR\,12567 (datasets 9 and 10), was observed more than once i.e. at different times during its emergence on the same day. We notice a decrease in the area covered by fast downflows with increasing age of this AR.

\subsection{Associated solar features}
\label{ssec:sd-loc}
The supersonic downflows are found in association with several solar features. In \tab{tab:loc}, we provide a classification of the fast downflow profiles based on the photospheric and chromospheric structures where they occur (column I). Pores, sunspot umbrae, sunspot penumbrae, and sunspot light bridges are identified using the continuum intensity images while the plage regions are identified with the help of \ion{Si}{I} Stokes $V$ maps. \ion{He}{I} loops (or AFS) and filaments (suspended above the polarity inversion line) are the chromospheric structures deduced from \ion{He}{I}c line center intensity images (see Figs.~\ref{fig:app-maps1}--\ref{fig:app-maps4}). The number of supersonic downflow profiles (${\rm N_{sd}}$) belonging to each of these solar features is given in column II. Column III shows what fraction of the total 13466 profiles displaying a supersonic downflow component is ascribed to a particular solar feature. Column IV gives the numbers of all datasets where flows in a given solar feature are identified.

As far as the photospheric structures are concerned, about 53\,\%\ of the locations displaying supersonic downflows are found in the periphery of pores (e.g. dataset 4), sunspot umbrae and penumbrae together account for 34\,\%\ (e.g. dataset 14) and about 3\,\%\ end up over light bridges (e.g. dataset 9). Slightly over 10\,\%\ of the supersonic downflows appear at seemingly quiet regions in the continuum intensity images, some of which are in immediate vicinity of ARs (e.g. dataset 1). The strong Stokes $V$ signals in the \ion{Si}{I} line at their locations indicate that they are AR plage. In the chromosphere, more than 91\,\%\ of the supersonic downflows are found in association with \he{} loops (as in dataset 8 where both footpoints of the loops are clearly visible or as in dataset 11 where only one footpoint of a loop is seen in the FOV). 12 out of the 14 datasets contribute to this category. These supersonic downflows lie along and at the footpoints of the loops, with the velocity increasing toward the footpoints. \he{} filaments (e.g. dataset 3) contribute to nearly 3\,\%\ of the observed supersonic downflows, and are mostly seen along the filament barbs, with increasing velocities away from the spine. A few supersonic downflows in seven of the datasets, which amount to the remaining 6\,\%, could not be clearly assigned to any structure and hence are grouped under the category "others". According to this classification, the majority of the fast downflows are located at the sites identified as pores and sunspots belonging to regions of emerging magnetic flux (EARs). This is in accordance with the inference that EARs host most of the fast downflows.

Moreover, a very crude calculation indicates that the total area covered by the supersonic downflow hosting pixels in pores are about 29.4\,\%\ of the total pixels covered by pores in 9 datasets. This fraction is about 1.7\,\%\ for the penumbrae in 9 datasets, 1.3\,\%\ for umbrae in 4 datasets, 8.2\,\%\ for light bridges in 2 datasets and 0.38\,\%\ for plages in 10 datasets. Further, for the \ion{He}{I} loops in 12 datasets and filaments in 2 datasets, these fractions turn out to be 2.46\,\% and 2.3\,\%, respectively. These estimates are biased since only those datasets which display supersonic downflows in a given category are used to determine the area covered by a given AR feature. Nevertheless, it is clear that supersonic downflows span only a tiny fraction of an AR feature's area, with the highest fraction being in pores.

\begin{table}[ht]
\centering
\caption{Classification of the supersonic downflows based on where they occur. The upper part of the table identifies photospheric features, the lower part chromospheric features. }
\begin{tabular}{m{3.3cm}rrr}
\midrule[1.5pt]
\multicolumn{1}{l}{Solar feature} & \multicolumn{1}{r}{N$_{\rm sd}$} & \multicolumn{1}{r}{\%} & \multicolumn{1}{r}{Dataset \#}\\
\midrule[1.5pt]
Pore (cont.) & 7173 & 53.27 & 1-4,6-8,10,13\\
Penumbra (cont.) & 4273 & 31.73 & 1,4,8-14\\
Umbra (cont.) & 279 & 2.07 & 7,8,10,14\\
Light bridge (cont.) & 360 & 2.67 & 5,9\\
Plage (\ion{Si}{I}) & 1381 & 10.26 & 1-4,6,8,11-14\\
\midrule[1.5pt]
Loops in AFS (\ion{He}{I}c) & 12319 & 91.48 & 1,2,4-11,13,14\\
Filament (\ion{He}{I}c) & 399 & 2.96 & 3,14\\
Others (\ion{He}{I}c) & 748 & 5.56 & 1,7,8,10,11,12,14\\
\midrule[1.5pt]
\end{tabular}
\label{tab:loc}
\end{table}
\begin{figure*}[htpb]
    \centering
    \includegraphics[scale=0.7,trim=0.0cm 1.5cm 0.0cm 0.1cm,clip]{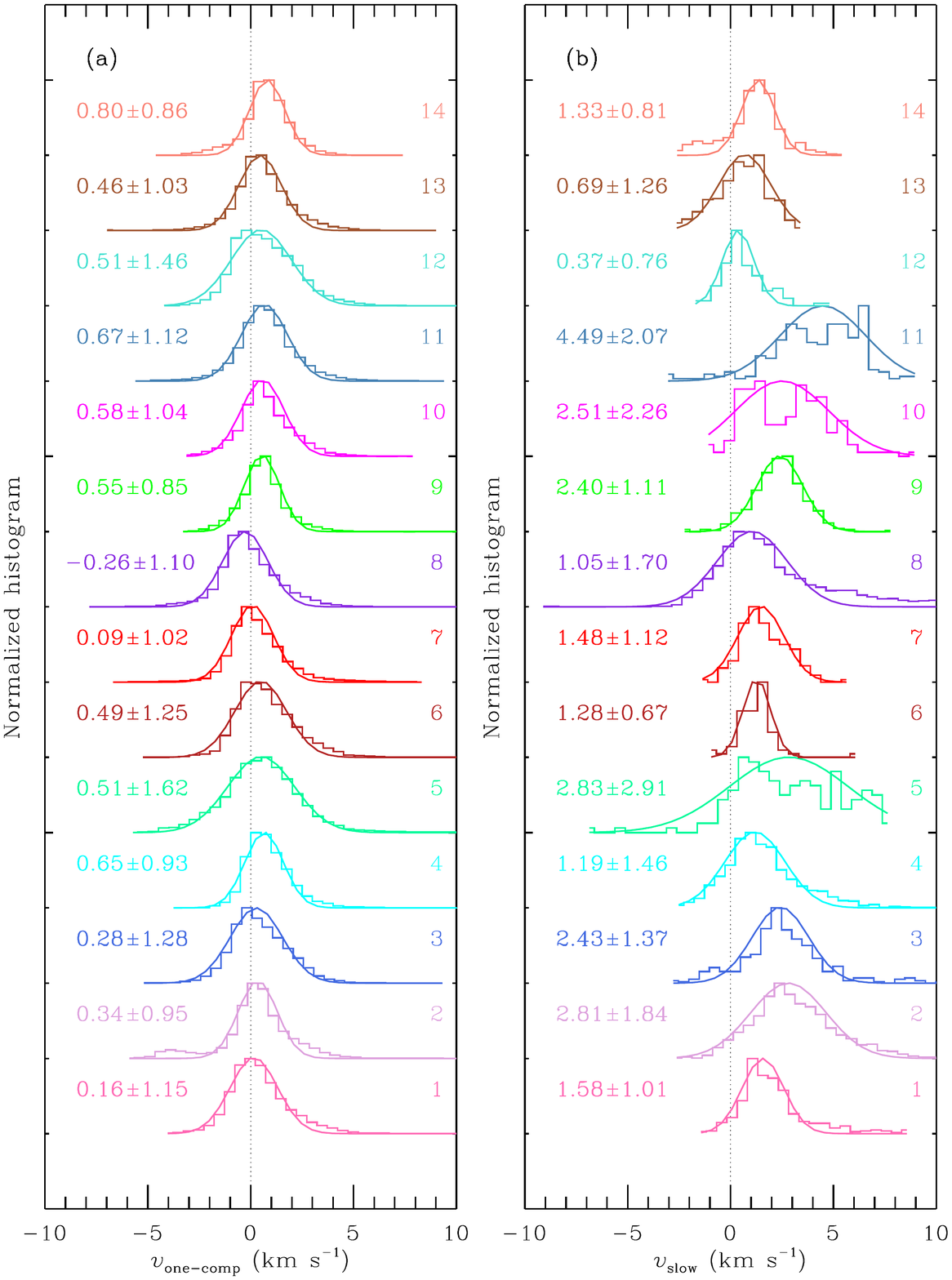}
    \caption{Panel (a): LOS velocity distribution of the subsonic flows at magnetized pixels ($M_S>2\sigma$ as discussed in \sref{ssec:class-obs}) with no supersonic downflows and the corresponding Gaussian fits. Panel (b): LOS velocity distribution of the subsonic component only for those pixels where supersonic downflow component coexists and the corresponding Gaussian fits. The numbers on the right in each panel represent the label of the dataset given in column I of \tab{tab:dataset}. The numbers on the left denote the mean and standard deviation of the velocity distributions, in \kms{}.}
    \label{fig:velc1}
\end{figure*}
\begin{figure*}[ht!]
    \centering
    \includegraphics[scale=.9,trim=1.5cm 7.5cm 0.cm 0.cm,clip]{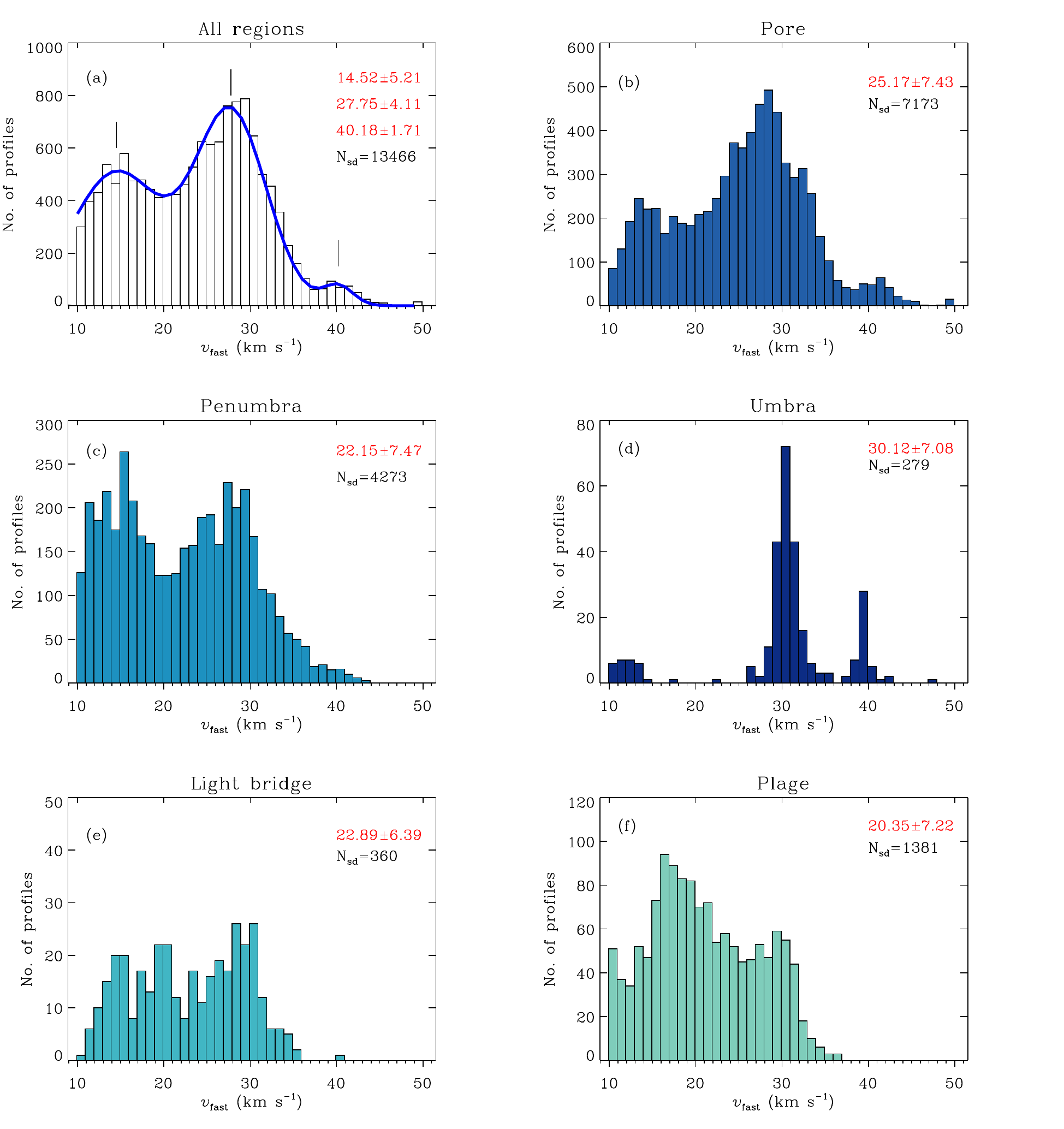}
    \caption{LOS velocity distribution of the magnetized supersonic downflows for all of the 14 datasets (panel a), pores (panel b), sunspot penumbrae (panel c), sunspot umbrae (panel d), light bridges (panel e) and plages (panel f). "N$_{\rm sd}$" is the number of supersonic downflow profiles in each category. The numbers in red are the mean and standard deviations of the distributions, in \kms{}. The blue curve in panel (a) is the triple-Gaussian fit to the velocity distribution. The vertical black solid lines mark the peaks of this triple distribution.}
    \label{fig:velc2}
\end{figure*}
\begin{figure*}[ht!]
    \centering
    \includegraphics[scale=0.9,trim=1.5cm 14.5cm 0.cm 0.cm,clip]{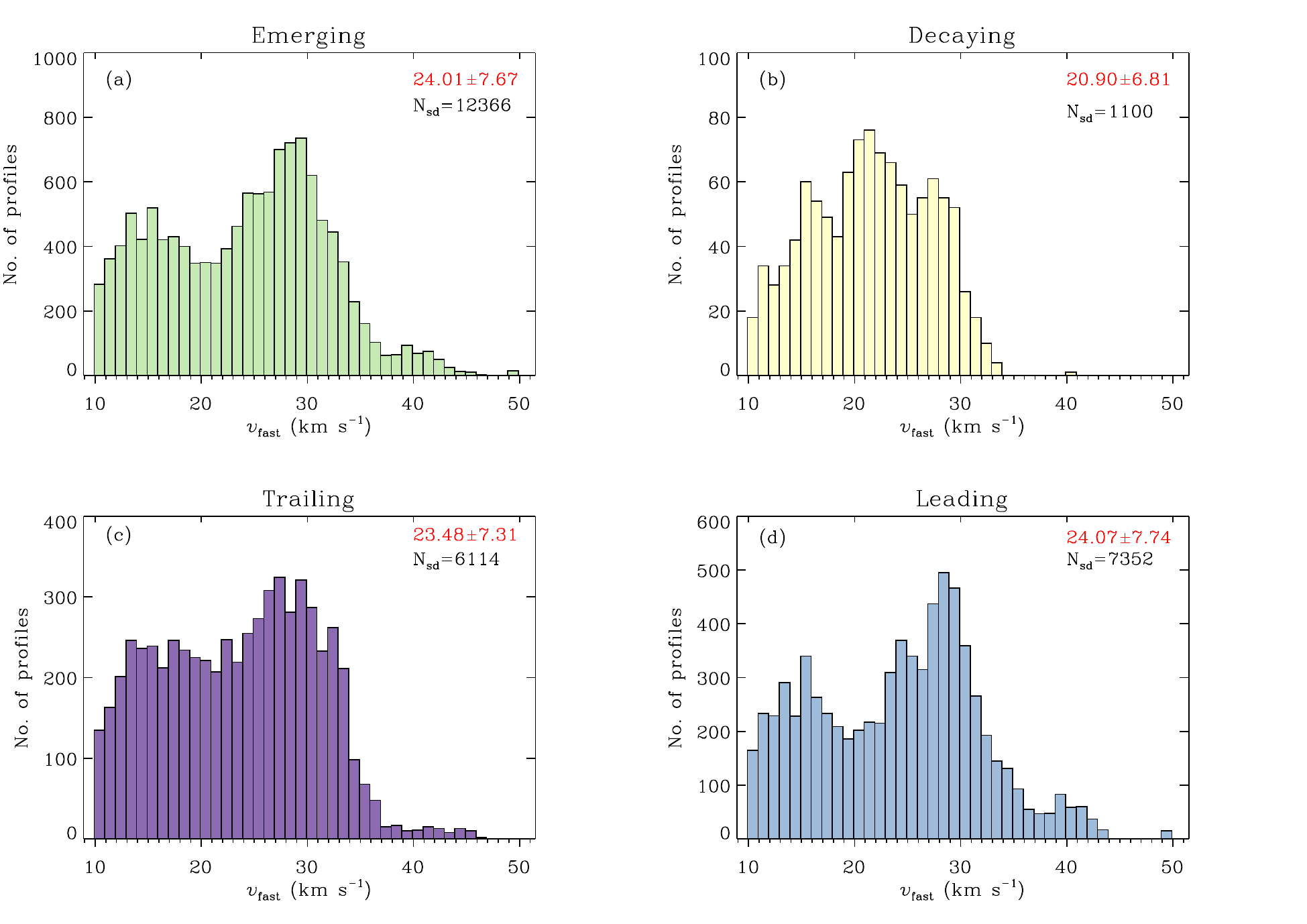}
    \caption{Distribution of the LOS velocities of supersonic downflows in EARs (panel a), DARs (panel b), trailing (panel c) and leading (panel d) groups (see \tab{tab:dataset}). The mean values and standard deviations of the distributions are given in red. "N$_{\rm sd}$" is the total number of profiles exhibiting supersonic downflows in each of the groups.}
    \label{fig:vel-dvse}
\end{figure*}
\begin{figure}[ht!]
    \centering
    \includegraphics[scale=0.5,trim=1.5cm 0.0cm 5.cm 15.cm,clip]{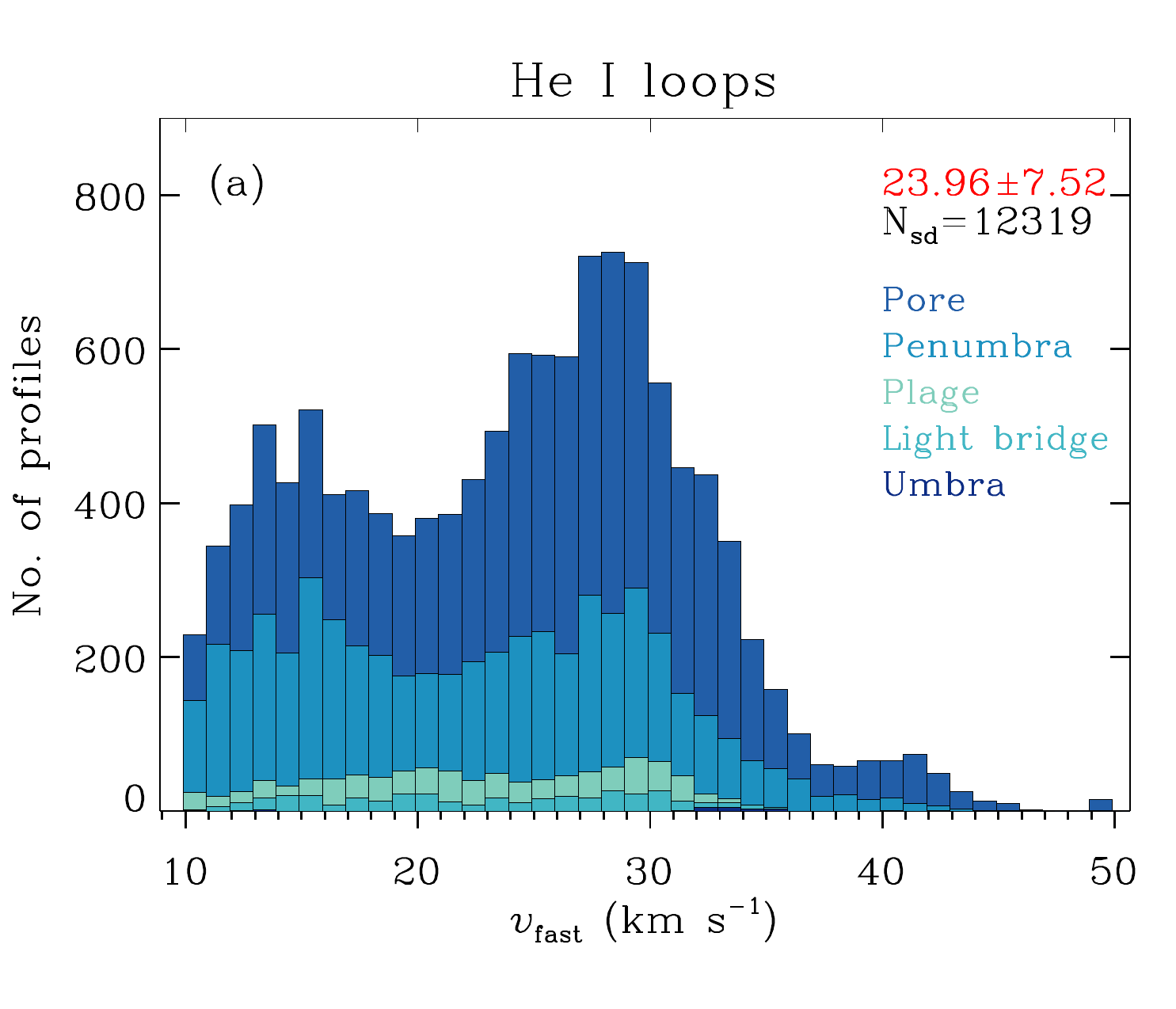}
    \includegraphics[scale=0.5,trim=1.5cm 0.0cm 5.cm 15.cm,clip]{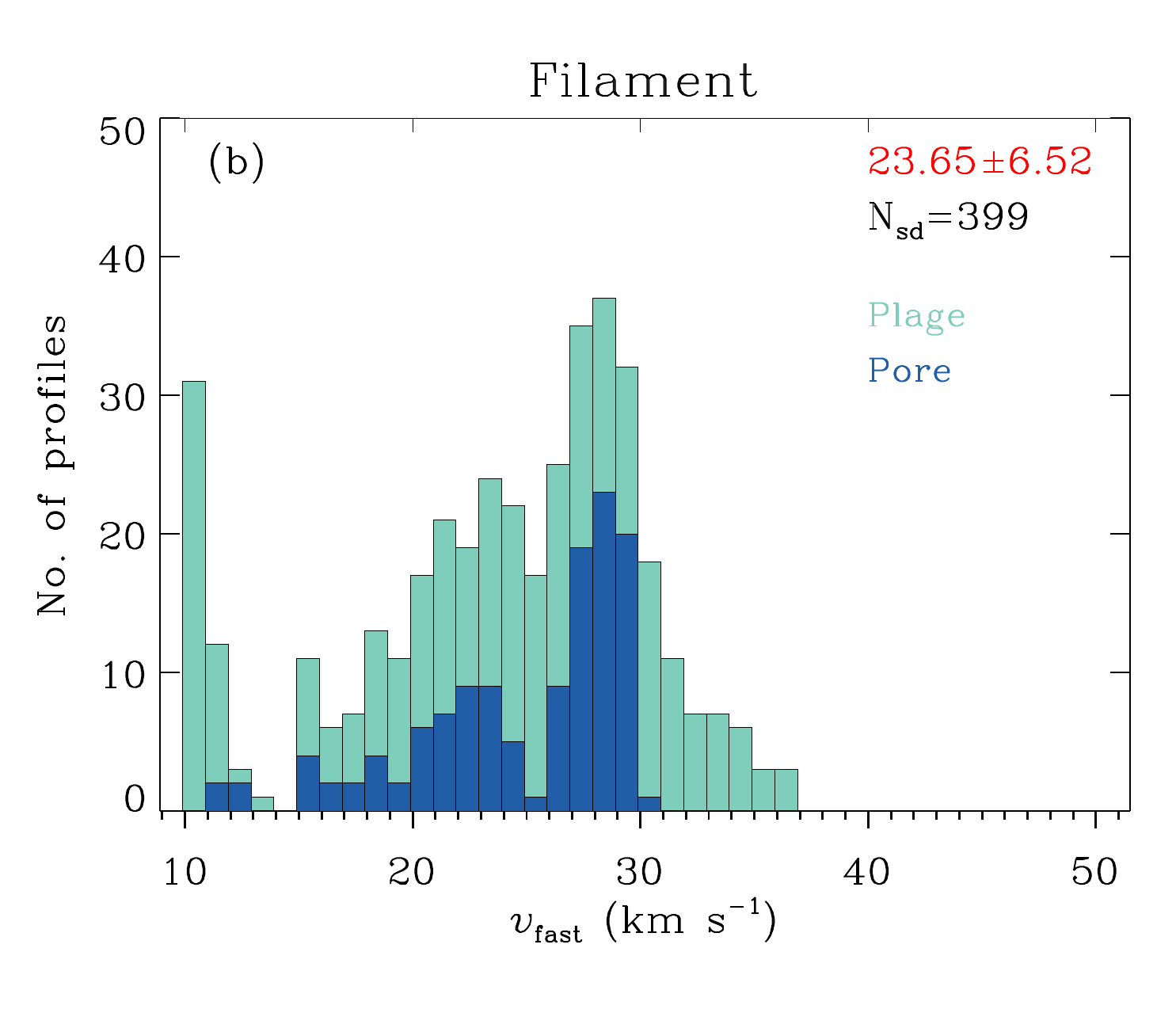}
    \includegraphics[scale=0.5,trim=1.5cm 0.0cm 5.cm 15.cm,clip]{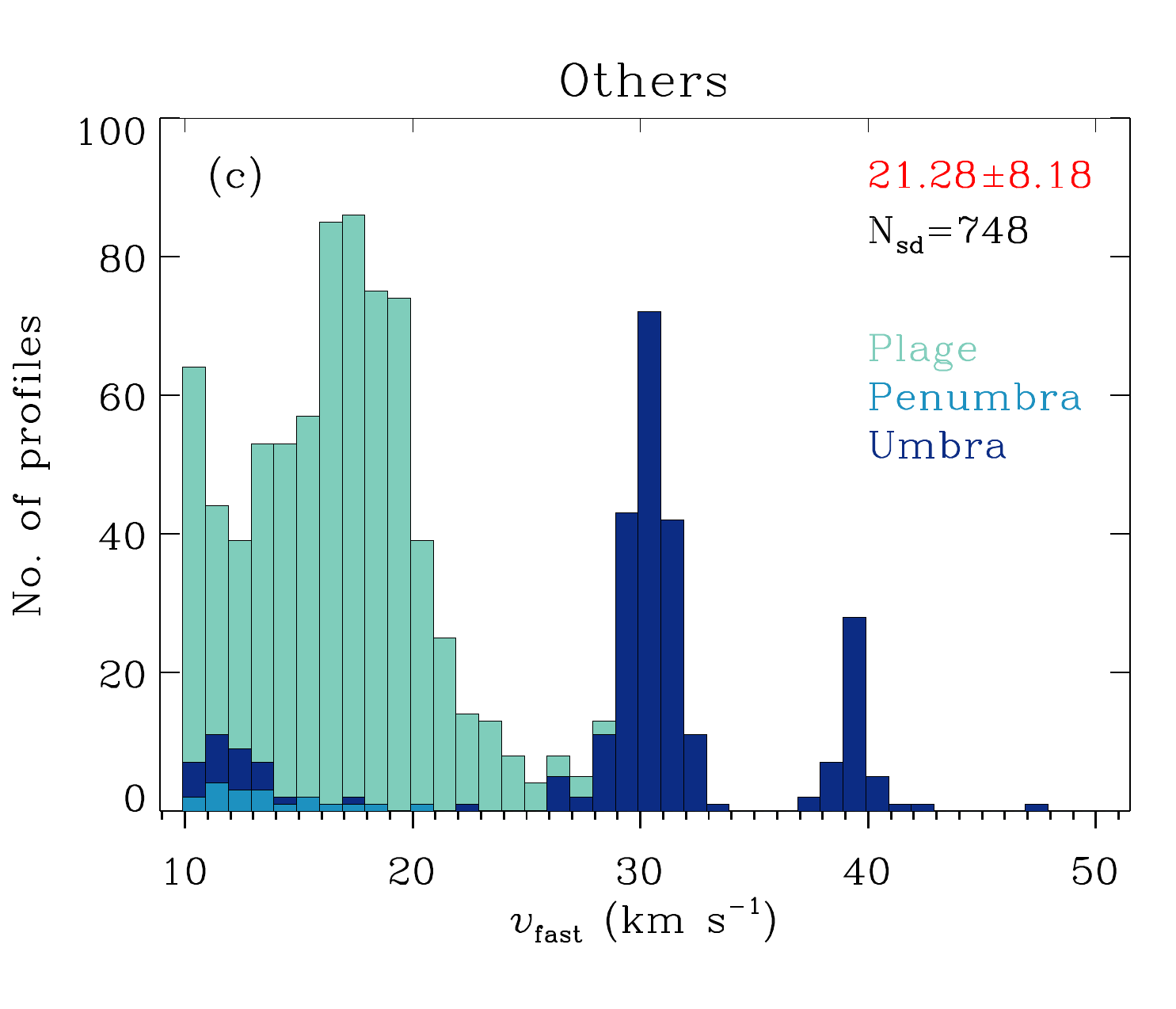}
    \caption{LOS velocity distribution of the magnetized supersonic downflows in He\,{\sc i} loops (panel a), filaments (panel b) and those for which the origin is not clear when considering the \ion{He}{Ic} line core images (panel c). The contribution of the photospheric features to each of the chromospheric category are shown. "N$_{\rm sd}$" is the number of supersonic downflow profiles in each category. The numbers in red are the mean and standard deviations of the distributions, in \kms{}.}
    \label{fig:velc3}
\end{figure}
\begin{figure*}[ht!]
    \centering
    \includegraphics[scale=.9,trim=1.5cm 9.cm 0.cm 0.cm,clip]{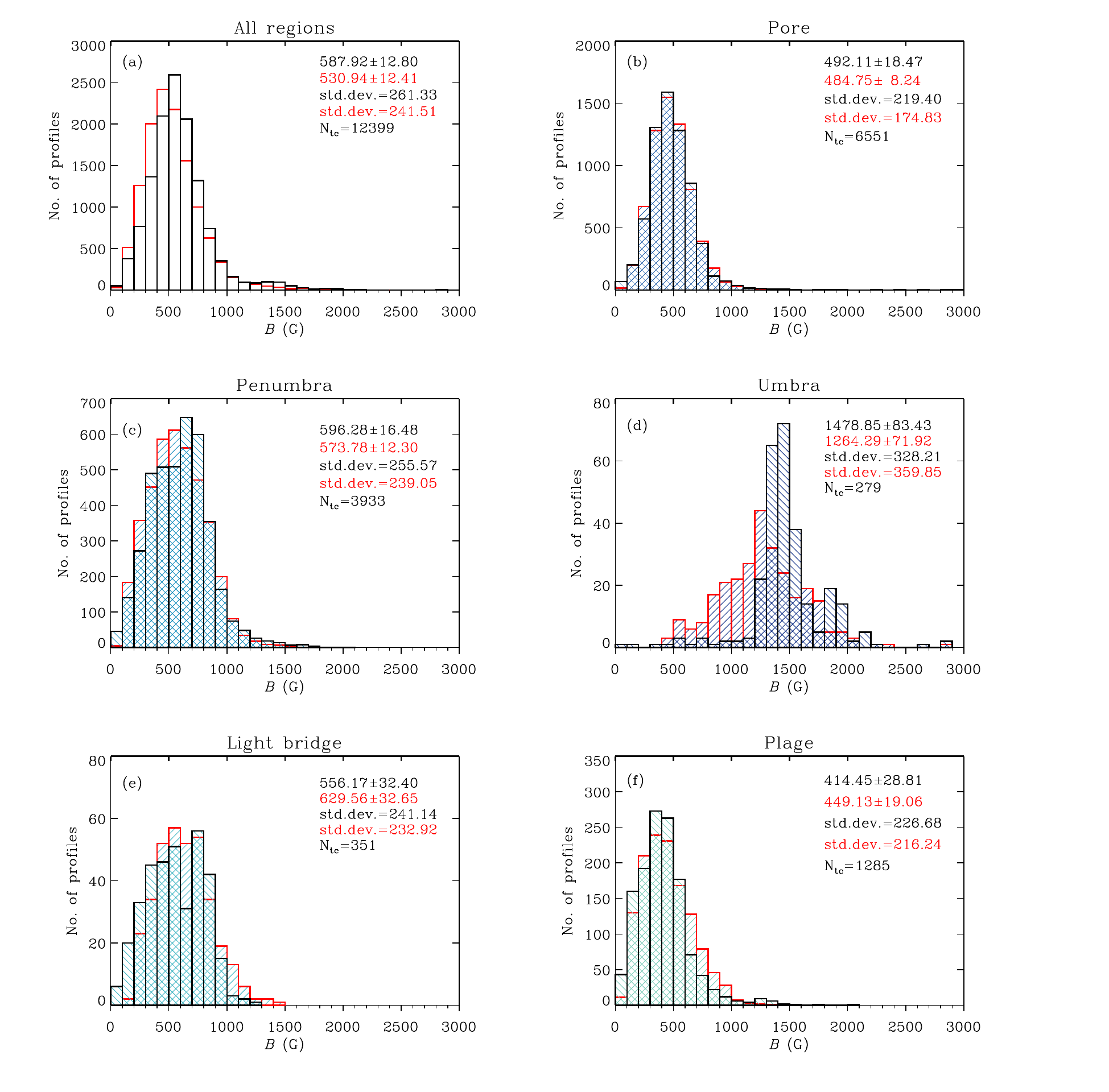}
    \caption{Magnetic field strength distributions of the slow (histograms outlined in black) and fast (histograms outlined in red) components for all of the 14 datasets (panel a), pores (panel b), sunspot penumbrae (panel c), sunspot umbrae (panel d), light bridges (panel e) and plages (panel f). Note only those profiles are counted that exhibit both slow and fast components. "N$_{\rm tc}$" is the number of profiles in each category where the supersonic downflow exists with a slow component. The numbers indicate the mean, uncertainty in the mean and standard deviations of distributions, in G, for slow (black) and fast (red) components.}
    \label{fig:bmagc2}
\end{figure*}
\begin{figure}[ht!]
    \centering
    \includegraphics[scale=0.5,trim=1.0cm 0.0cm 5.cm 15.cm,clip]{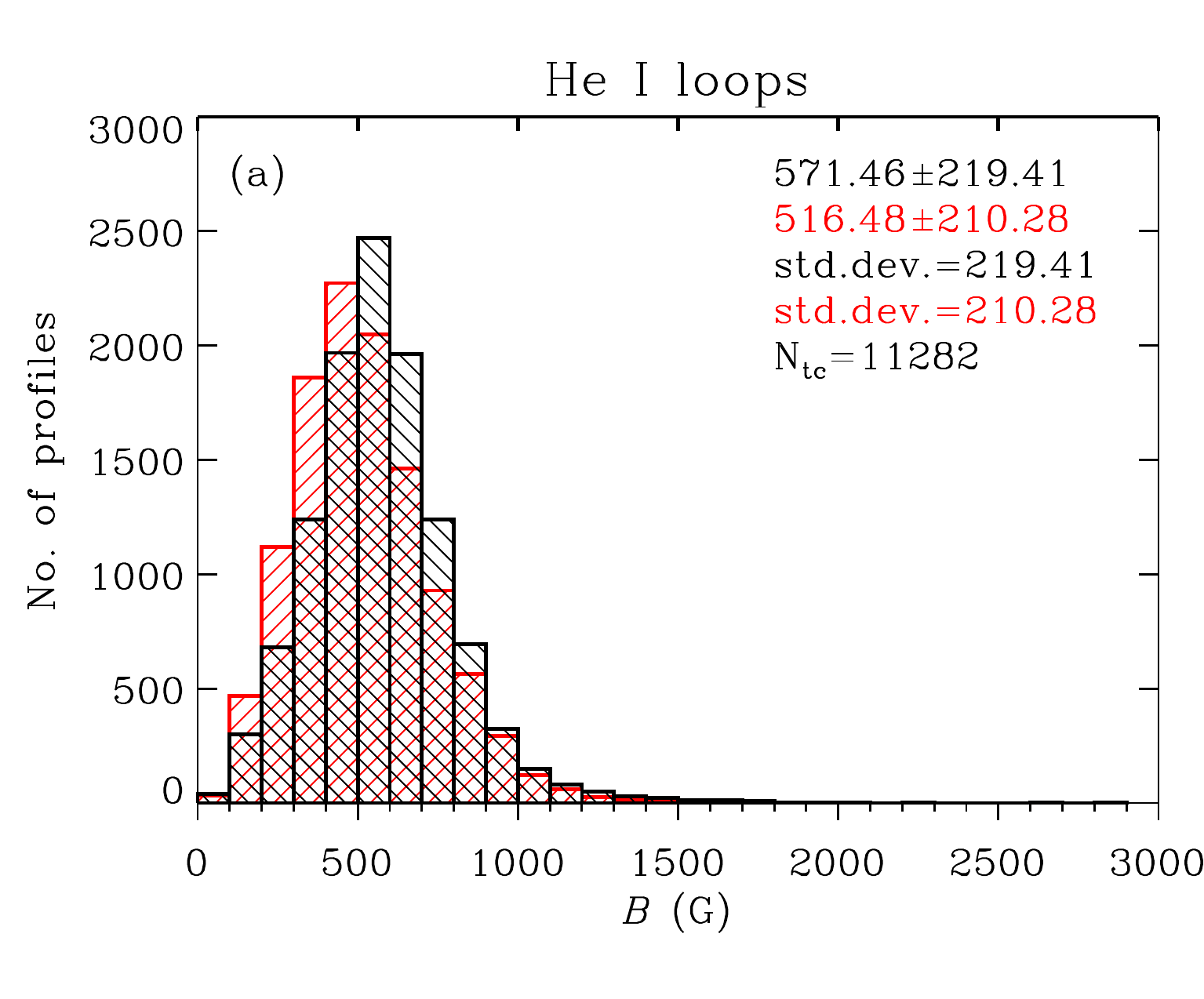}
    \includegraphics[scale=0.5,trim=1.0cm 0.0cm 5.cm 15.cm,clip]{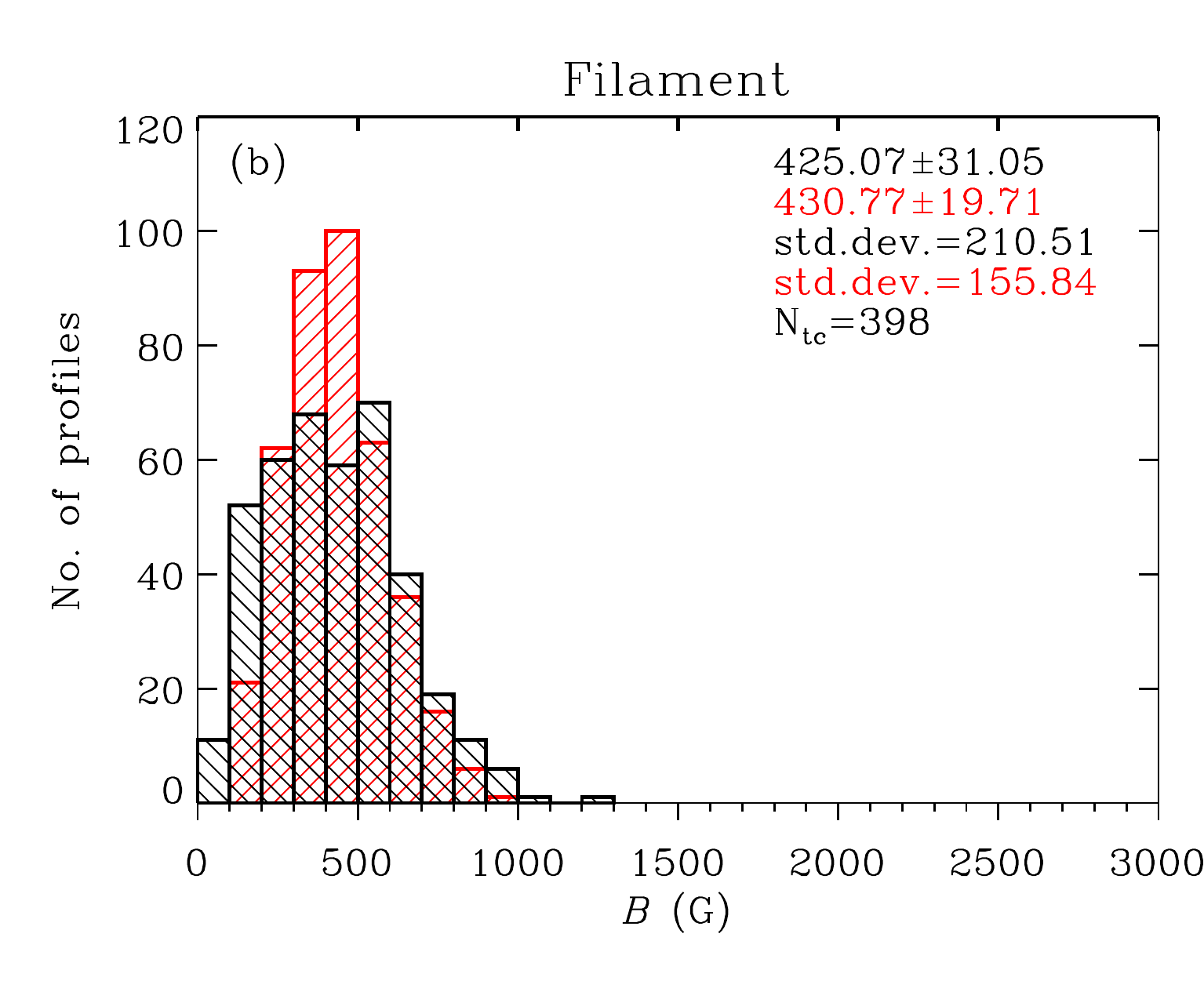}
    \includegraphics[scale=0.5,trim=1.0cm 0.0cm 5.cm 15.cm,clip]{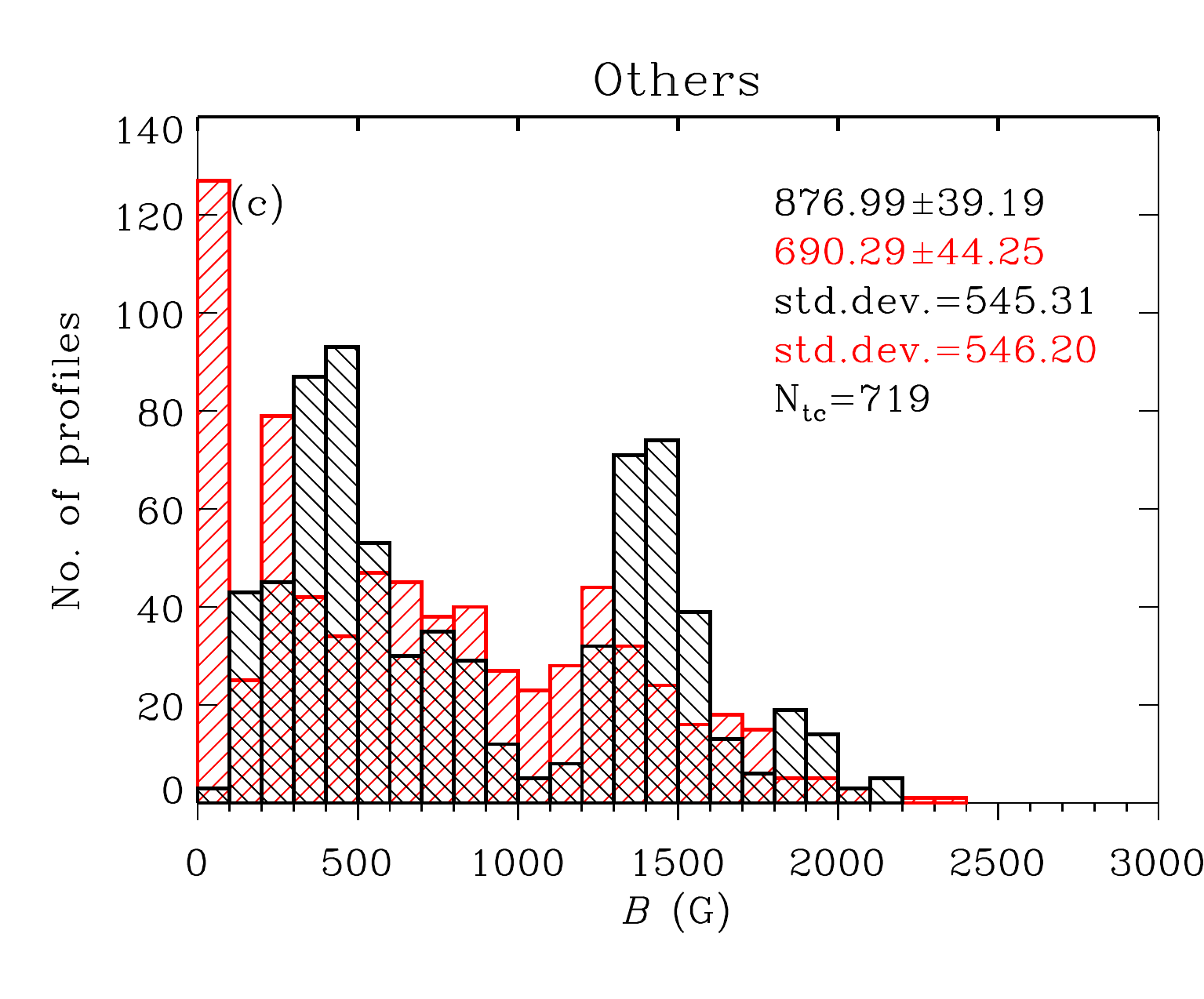}
    \caption{Chromospheric magnetic field strength distributions of the slow (black) and fast (red) components in He\,{\sc i} loops (panel a), filaments (panel b), and in chromospheric features that could not be clearly identified (panel c). Note only those profiles are counted that exhibit both slow and fast components. "N$_{\rm tc}$" is the number of profiles in each category where the supersonic downflow exists with a slow component. The numbers indicate the mean, uncertainty in the mean and standard deviations of distributions, in G, for slow (black) and fast (red) components.}
    \label{fig:bmagc3}
\end{figure}

\subsection{LOS velocity distribution}
\label{ssec:sd-vel}
The LOS velocity maps for the slow component are shown in Figs.~\ref{fig:app-maps1}--\ref{fig:app-maps4} (see also \fig{fig:invmaps}). The maps have been smoothed with a median filter for better representation. One of the apparent features that one can notice are the alternate red and blueshifts in the LOS velocity maps for the slow component in sunspots, for instance, as in datasets 1, 12, 13, and 14. These velocity oscillations (captured by the temporal scanning) are visible both in the umbra and penumbra \citep[see e.g.][]{2007ApJ...671.1005B}. We note that such velocity oscillations are not seen in the fast component. Such a case where the slow component shows oscillations while the fast component does not, was discussed by \citet{2001ApJ...552L..77B} for flows in the sunspot transition region (TR) which led them to suggest that the two components originate at different heights in the solar atmosphere (see further discussions in \sref{ssec:sd-orig}).

Figure~\ref{fig:velc1}a shows the normalized histograms of the LOS velocities of the subsonic flows at pixels classified as magnetized (column IV of \tab{tab:pixel-wlbin}) which do not host supersonic downflows. The distributions in \fig{fig:velc1}a are well fit by the Gaussian functions suggesting that the LOS velocities follow a normal distribution. However, the distributions are not always centered at 0\,\kms{} but is often shifted to the red (by up to 0.8\,\kms{}), as determined from the offsets of the best-fit Gaussians. Figure~\ref{fig:velc1}b shows the velocity distribution for the subsonic flows at only those pixels also hosting supersonic downflows. The Gaussian functions fit the velocity distributions fairly well for most datasets. Further, \fig{fig:velc1}b shows that the subsonic components of dual flows are predominantly downflows, with most of the Gaussians more redshifted than in \fig{fig:velc1}a.

The locations of occurrence of supersonic downflows fulfilling the selection criteria defined in \sref{ssec:criteria} are overplotted on the intensity images in Figs.~\ref{fig:app-maps1}--\ref{fig:app-maps4} (see also \fig{fig:invmaps}). The downflow speeds are given by green (for speeds up to 20\,\kms{}) and red shading (for higher speeds). The supersonic downflow velocities averaged over the individual datasets range between $15-27$\,\kms{}, while the maximum downflow velocities attained in the different datasets are in the range $22-49$\,\kms{} (see \tab{tab:pixel-wlbin}).

In order to check if the supersonic downflows show different characteristics according to the solar features they are associated with, we plotted their velocity distributions for the different AR structures given in \tab{tab:loc}. Such distributions for the photospheric structures are shown in \fig{fig:velc2}. In panel (a), we have shown the velocity distribution of all 13466 profiles combined from the 14 datasets. In this case, the LOS velocities present a distinct double-peaked distribution and a less prominent third peak at higher velocities. The velocity distribution in \fig{fig:velc2}a shows that our choice of 50\,\kms{} as an upper limit for the fast component velocities is reasonable as none of the flows reach this boundary. By fitting three Gaussians to these distributions (the blue curve), we determine the peak velocities to be 14, 28, and 40\,\kms{}. For a better visualization of the locations of the two big populations, in Figs.~\ref{fig:app-maps1}--\ref{fig:app-maps4}, we have indicated all supersonic downflows with velocities up to 20\,\kms{} by green shaded symbols while the remaining supersonic downflows are shown by red shaded symbols. The small third population is counted to the second population in this case.

Figure~\ref{fig:velc2}b shows the velocity distribution of the supersonic downflows located in the periphery of pores. More than half of the observed supersonic flows lie in this category and are mostly constituted by the large fraction of downflows in datasets 2, 4 and 8. Dataset 4 displays the fastest of all the observed supersonic downflows in our sample with LOS velocities up to 49.2\,\kms{} (see column VII of \tab{tab:pixel-wlbin}). The overall distribution of the LOS velocities appears very similar to that in \fig{fig:velc2}a. The velocities of the supersonic flows ending in penumbrae (a great part of which comes from dataset 8 alone) also show a distribution similar to those in pores, however with the first and the second populations being equally strong (\fig{fig:velc2}c). 

The distribution of the supersonic downflows found above sunspot umbrae appears distinct from that of penumbrae. The majority of these flows are faster than 20\,\kms{}, with a slightly higher average of $\sim30$\,\kms{}. The first group of profiles with velocities below 20\,\kms{} is constituted by the downflows in datasets 8 and 14. The second group around 30\,\kms{} is mostly due to dataset 14 while the third group at 40\,\kms{} is dominated by dataset 10. The downflows above light bridges resemble those in penumbrae and have similar average velocities. However, the flow speeds on the whole lie below 40\,\kms{}, which is also the case for the downflows in category plage. It is clear from these distributions that the downflows in light bridges and plage do not contribute to the population peaking at 40\,\kms{} in \fig{fig:velc2}a.

When the downflow profiles are grouped into those from EARS (consisting of 9 datasets; \fig{fig:vel-dvse}a) and DARs (consisting of 5 datasets; \fig{fig:vel-dvse}b), we obtain a distribution with multiple populations which resembles that of \fig{fig:velc2}a for EARs while for the DARs, the distribution appears to consist of a single population with continuous LOS velocity values. To check if there is any dependence of the downflow velocities on the leading or trailing part of the ARs, we classified the observed regions into leading and trailing groups using the LOS magnetograms from HMI (\tab{tab:dataset}, column XII). There are two datasets containing both the leading and trailing polarities. For these cases, we separated the profiles from trailing and leading polarities using the LOS magnetic field strength maps ($B\cos\gamma$) created from our inversion code output. The velocity distribution for the leading and trailing groups are shown in panels (c) and (d) of \fig{fig:vel-dvse}. In both groups, the distributions peak at nearly the same velocities, suggesting that the multiple populations in the downflow velocities are independent of whether the downflows are found in the leading or trailing side of the AR.

Figure~\ref{fig:velc3} shows the LOS velocity distribution of downflows in chromospheric features. The distributions are color coded to highlight the contribution from various photospheric categories to each of the chromospheric ones. The distribution for \ion{He}{I} loops (the most common site for supersonic downflows) in \fig{fig:velc3}a is similar to that for pores, with an average velocity close to 24\,\kms{}. Filaments host supersonic downflows whose LOS velocities show a nearly continuous distribution \citep[see e.g.][]{2020IAUS..354..454S}. A set of profiles in \fig{fig:velc3}b which have speeds below 15\,\kms{} arise from a part of the filament seen in dataset 14. The distribution of downflows in the category "others" (\fig{fig:velc3}c) for which the associated structure was unclear if we consider just images taken in the core of the \ion{He}{Ic} line, appear to be a combination of the distributions seen for plage and umbrae. Dataset 12, that shows downflows in this category was previously studied by \citet{2018A&A...614A...2V}. The locations of the supersonic downflows we derived are consistent with the locations where \citet{2018A&A...614A...2V} reported strong downflows in \ion{He}{I} using a single Lorentzian fit.

Finally, we note that to our knowledge this study forms the first comprehensive statistical analysis of the chromospheric magnetized supersonic downflows. However, since the statistics are small in the sense that relatively few ARs were analyzed, the relative number of supersonic downflows in each category discussed above should be taken with caution.

\begin{figure}
    \centering
    \includegraphics[scale=0.5,trim=1.0cm 0.0cm 5.cm 15.cm,clip]{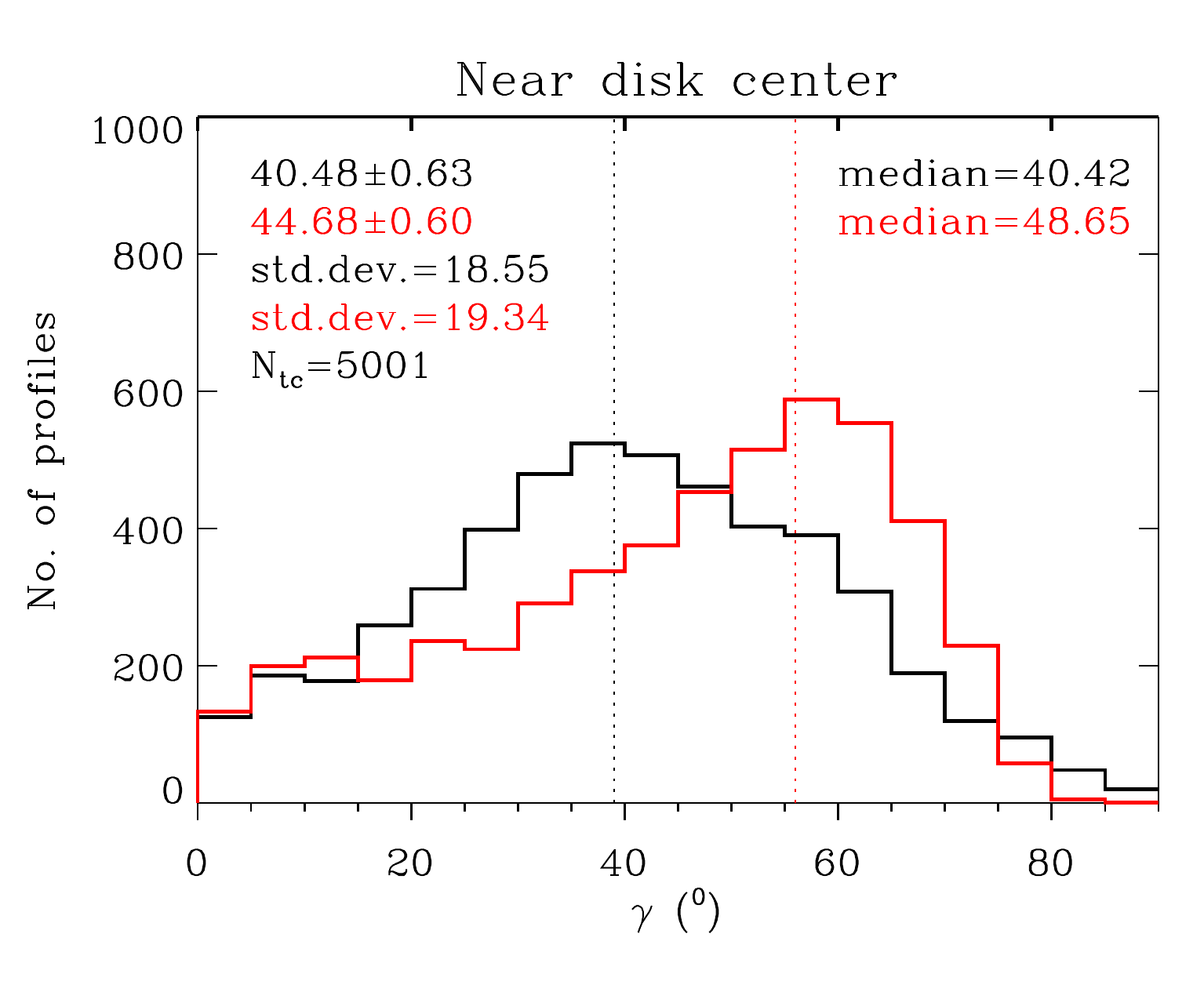}
    \caption{Distributions of the inclination of the magnetic field for slow (black) and fast (red) components for all datasets observed closer to the disk center ($\mu>0.9$). "N$_{\rm tc}$" is the number of profiles in each category where the supersonic downflow exists with a slow component. The numbers indicate the mean, the uncertainty in the mean, standard deviations and the median values of distributions for slow (black) and fast (red) components. The vertical dotted lines mark the mode values of the distributions.}
    \label{fig:inclc2}
\end{figure}

\subsection{Magnetic field}
\label{ssec:sd-mag}
In the regions where supersonic downflows occur, \he{} lines mostly show polarization signals indicative of the Zeeman effect suggesting that the magnetic field strengths in those regions are beyond the Hanle sensitivity regime. However, in datasets 7 and 8, polarization signals in $Q$ and $U$ depicting the modification of scattering polarization by the magnetic field via the Hanle effect \citep[see e.g.][for a discussion of the Hanle effect in \he{} triplet]{2002Natur.415..403T,2004A&A...414.1109L,2008ApJ...683..542A} are noticeable at a very few pixels which harbour supersonic downflows close to the footpoints of \ion{He}{I} loops. In our sample of 13 AR scans, the linear polarization signals due to Hanle effect are also found in pixels which do not host supersonic downflows (notably in \ion{He}{I} loop tops). We remark that the Hanle effect is not included in our inversions and the few profiles showing Hanle effect are still treated under the approximation of the Zeeman effect. The fraction of supersonic downflow-hosting pixels that also exhibit Hanle signatures is less than 1\,\%\ of the total number of pixels with supersonic downflows. Therefore the influence of this approximation on the statistical results presented here is negligible.

In Figure~\ref{fig:bmagc2}, we show the distributions of the magnetic field strength for photospheric structures at only those pixels where both, supersonic downflows (red histograms) and subsonic up/downflows (black histograms) coexist. There are in total 12399 such pixels. Panels (a)--(f) show these histograms for the various solar structures indicated in the titles of these panels. N$_{\rm tc}$ gives the total number of two-component profiles in each category. The number of profiles in each panel is reduced from the corresponding ones in \fig{fig:velc2} as we have excluded the 1067 profiles showing only fast downflow with supersonic velocities (i.e., without a coexisting slow component). The mean value of the field strength and standard deviation of the distribution are given in black for the slow component and in red for the fast component. When all the 12399 profiles are considered together (panel a), on average the magnetic field has a strength of about 530\,G in the fast component whereas it is stronger by approximately 50\,G in the slow component. The distribution, however, is similar in the two components. In the pore vicinity, penumbrae and umbrae, the subsonic component is associated with somewhat stronger fields (by $10-200$\,G on average). On the whole, the result that the slow and fast components have different field strengths is consistent with the findings of \citet{2007A&A...462.1147L}. Chromospheres above sunspot umbrae have the highest mean field strength of all the analyzed structures, showing 2\,kG at the locations of supersonic downflows. In light bridges and plage regions, however, the supersonic component displays slightly stronger magnetic fields.

The magnetic field strength distributions of the two components in various chromospheric structures are shown in \fig{fig:bmagc3}. The characteristics of the distributions in \ion{He}{I} loops are similar to those in Figure~\ref{fig:bmagc2}a. Filaments show magnetic fields which are nearly equally strong in the two components, although the field strength associated with the supersonic flows are generally somewhat weaker. The strongest fields (essentially due to the umbrae) are established in the category where it was not possible to decipher the chromospheric structure hosting the supersonic downflows.

For an accurate determination of the inclination of the magnetic field vector, the linear polarization signals should stay well above the noise level. We find that the majority of the profiles associated with strong downflows exhibit very weak linear polarization signals which lie below the noise threshold. To determine supersonic downflows discussed here we set the requirement that one of the three Stokes polarization parameters is sufficiently above the noise. To not bias the results too much by excluding a large number of spatial pixels, we set the same requirement for determining the inclination. We stress, however, that as a consequence of this, the inclination values should be taken with caution, as they may be overestimated (since Stokes $Q$ and $U$ tend to be more strongly affected by noise - \citealt{2011A&A...527A..29B}).

In \fig{fig:inclc2}, we show the distribution of the inclination for the slow (black) and fast (red) components for regions observed close to the disk center ($\mu>0.9$). The LOS inclinations retrieved from the inversions are closer to the true inclination of the magnetic field when the regions are observed close to the disk center. For regions observed away from the disk center, the interpretation of the field inclination is not straightforward without the LOS azimuth disambiguation. Hence we restrict the analysis of magnetic field inclination only to the near disk center cases. In the inversions, the inclination takes values in the range between 0\textdegree\ -- 90\textdegree\ for positive polarity and between 90\textdegree\ -- 180\textdegree\ for the negative polarity. For the purpose of \fig{fig:inclc2}, the inclination values between 90\textdegree\ -- 180\textdegree\ have been folded on to the range 90\textdegree\ -- 0\textdegree. For the slow component, about 59\,\%\ of 5001 profiles give inclination below 45\textdegree\ (which is already evident from the mean and median values of the distributions). This suggests that the magnetic field is more vertical with respect to the solar surface in the regions where the slow component originates. The magnetic fields associated with the supersonic downflows appear to be more horizontal. About 56\,\%\ of the 5001 profiles result in inclination values higher than 45\textdegree\ (with a median value of $\sim$48\textdegree).  This difference in the inclination distribution for the slow and the fast components, is consistent with the findings of \citet{2007A&A...462.1147L}.

In their studies with the quiet Sun data, \citet{2011A&A...527A..29B,2012A&A...547A..89B} pointed out that inversions tend to retrieve more horizontal fields when the polarization signals are weak, owing to the influence of photon noise. To determine if the more horizontal field inclinations retrieved for the fast component is due to weaker magnetic signals in them compared to the ones in the slow component, we did a test. For this we selected all pixels where the two components had similar filling factors and checked the distributions of the field inclination for the two components. Even when the magnetic signals are equally strong in the two components, we found that the field for the fast component is more horizontal than for the slow component. All in all, the magnetic field strength and its inclination for the slow and fast components suggest that the fast component is formed higher up, for flows occurring along a leg of a loop.

\section{Discussion}
\label{sec:sd-disc}
\subsection{Redshifted velocities}
\label{ssec:sd-redshift}
At the formation temperature of about 8000--10000\,K of the \he{} triplet, the sound speed is around 10\,\kms{}. As listed in \tab{tab:pixel-wlbin}, the average velocities of the observed magnetized supersonic downflows are $1.5-2.7$ times the local sound speed \citep[in accordance with the values given in][]{2020ApJ...890...82G}. The maximum velocity of these downflows exceeds the sound speed by at least 2 times and reaches values as high as 49.2\,\kms{}. This is in agreement with the findings of \citet{1997SoPh..172..103M,1998ASPC..155..341M,2000ApJ...544..567S,2007A&A...462.1147L} and more recently \citet{2018A&A...617A..55G}, who have all observed velocity redshifts of $\sim40$\,\kms{}. In some of the earlier studies, even larger redshifts have been reported, although these were typically associated with flares. For example, in a \he{} filament observed during the decay phase of a flare, \citet{1995ApJ...441L..51P} found high redshifts of about 60\,\kms{}. In a flaring AR filament, \citet{2011A&A...526A..42S} observed redshifts of up to 100\,\kms{}. However, in the ARs that we analyzed, none of which was flaring at the time of the observations, no downflows faster than 50\,\kms{} were seen.

The magnetized supersonic downflows associated with pores, penumbrae, and \ion{He}{I} loops show multiple populations in their LOS velocity distribution with peaks roughly around 14, 28, and 40\,\kms{} as shown in Figs.~\ref{fig:velc2}~and~\ref{fig:velc3}. Note that these pores and penumbrae are located at the footpoints of the \ion{He}{I} loops. The wavelength separation between the strong telluric line at 10832.09\,\AA{} and the \ion{He}{Ic} line at 10830.34\,\AA{} corresponds to a velocity shift of $\sim48$\,\kms{}. The peaks of all the three populations in LOS velocities are well below 48\,\kms{} indicating that none of the three populations are related to the telluric blend line (we found a cluster of downflows at that velocity before we decided to concentrate on magnetized downflows). In particular, the third, i.e. fastest population is not seen in the velocity distribution for features like light bridge, plage, and filaments and hence cannot be arising from the telluric blend. The magnetic nature of the supersonic downflows analyzed ensures that the artifacts due to telluric line are removed, since the telluric line does not have any magnetic signal of its own.

A possible explanation for the broad range of downflow velocities seen for the AFS loops might come from the differences in the heights of the loops hosting supersonic downflows. The ARs showing AFS loops are likely at different phases of emergence, with loops on average at different heights. The gravitational acceleration results in this case in different velocities at the footpoints with high lying loops showing higher velocities. See \citet{2018A&A...617A..55G} for a description of the temporal evolution of the LOS velocities in an arch filament during its lifetime. However, it remains unclear why there are three distinct ppopulations. Lastly, we remark that such multiple populations in fast downflow velocities were also reported in \citet{2007msfa.conf..173A}, who analysed 35 scans of 13 ARs recorded using the German Vacuum Tower Telescope and determined the LOS velocities using the multiGaussian fit to the intensity profiles. They found that the \he{} supersonic downflows have two distinct populations: the first population had LOS velocities up to 17\,\kms{} and peaked at around 10\,\kms{} while the second population peaked around $20-25$\,\kms{}.

\subsection{Origin of magnetized supersonic downflows}
\label{ssec:sd-orig}
Several scenarios have been put forward in the literature to explain the origin of highly redshifted flows. Here, we discuss the possible mechanisms behind the supersonic downflows that we observed in various AR features.

Many datasets in our sample show rapid chromospheric downflows in the vicinity of pores located at the footpoints of \ion{He}{I} loops, similar to the ones studied by \citet{2007A&A...462.1147L}. In fact, the fastest downflows are seen around pores i.e. relatively cool structures harbouring strong magnetic fields. The lower temperature leads to a reduction in the pressure scale height and reduces the gas pressure at a given height. Increased magnetic field concentration leads to an enhancement in the Wilson depression. Following \citet{2007A&A...462.1147L}, the association of fast downflows with pores could be explained by the small pressure scale height, which leads to a stronger evacuation of the flux tube at the footpoint of the loop, at least for heights at which the temperature is lower than in the surroundings. Because of this the cooler material carried by the \ion{He}{I} loop can travel down to a larger distance before it hits gas at a given density or pressure, attaining higher velocities in the process. If the pore is forming and the field is increasing, then the convective collapse of the photospheric flux tube removes the hydrostatic support of the chromospheric plasma, causing the gas in the photospheric layers of the pore to flow down as well \citep[e.g.][]{1978ApJ...221..368P,1979SoPh...61..363S,1998A&A...337..928G}.

Supersonic downflows seen above sunspots could be the chromospheric counterparts of TR supersonic downflows. Multicomponent flows within a resolution element are a well-known observational features of TR spectral lines. Localized supersonic downflows above sunspots were in fact first seen in the TR spectral lines observed with the High-Resolution Telescope and Spectrograph by \citet{1982SoPh...77...77D,1988ApJ...334.1066K,1990Ap&SS.170..135B,1993SoPh..145..257K}. These authors ascribed such multiple flows to the fine structure of the TR i.e. to the presence of multiple fibrils within a given resolution element. In the observations by the Solar Ultraviolet Measurements of Emitted Radiation instrument \citep[SUMER;][]{1995SoPh..162..189W} onboard the Solar and Heliospheric Observatory (SOHO), \citet{2001ApJ...552L..77B} found dual flows. The low speed component showed clear association with the umbral oscillations while the faster one did not, suggesting that the two components are formed at different heights along the LOS \citep[see also][]{2004ApJ...612.1193B}. The data from the Interface Region Imaging Spectrograph (IRIS) also revealed the presence of steady \citep[e.g.][]{2015A&A...582A.116S} as well as bursty, shot-lived \citep[e.g.][]{2014ApJ...789L..42K} supersonic downflows above sunspots \citep[see also][]{2020A&A...636A..35N}. \citet{2018ApJ...859..158S} carried out a statistical analysis of the TR supersonic downflows in 60 sunspot datasets observed by IRIS and found that in more than 40\,\%\ of the cases, the TR supersonic downflows exhibited chromospheric supersonic downflows. This points to a possibility that some spots in our sample probably show the chromospheric equivalents of TR supersonic downflows. It is also possible that some of the downflows that we observe are associated with sunspot plumes \citep{2001SoPh..198...89B,2001ApJ...552L..77B}.

Evershed effect refers to the horizontally outward flow of gas observed in the penumbral filaments at the photosphere \citep{1909MNRAS..69..454E}. A reversed flow called the inverse Evershed flow is seen in the chromosphere \citep[][and references therein]{1975SoPh...43...91M,2003A&ARv..11..153S}. For the sunspot observed in dataset 1, most of the supersonic downflows are on the side of the spot showing redshifts due to the inverse Evershed flow (see LOS velocity map in \fig{fig:app-maps1}). 

\citet{2009ApJ...704L..29L} reported photospheric supersonic downflows in a sunspot light bridge using measurements taken with the spectropolarimeter onboard the {\it Hinode} satellite. They found the supersonic downflows to be located at regions where the magnetic field of the light bridge meets the sunspot field having a different orientation, and to be co-spatial with the brightenings seen in the chromospheric \ion{Ca}{II} H filtergrams. They speculated these supersonic downflows to be the result of the magnetic reconnection happening in the upper photosphere/lower chromosphere. In two datasets in our sample, supersonic downflows are seen above sunspot light bridges, many of which seem to be located near the footpoints of the \ion{He}{I} loops anchored in the light bridges. The flow velocity appears to increase toward the footpoints, indicating plasma motion along the loops, similar to that seen for pores. Further, brightenings in the corona as sampled by the 171\,\AA{} channel of the Atmospheric Imaging Assembly (AIA) onboard SDO, which could be the signatures of magnetic reconnection, are seen in the case of dataset 5. Hence, the possibility that the small fraction of supersonic downflows above light bridge which are not associated with \ion{He}{I} loops arise from magnetic reconnection in the corona cannot be ruled out.

\citet{2000ApJ...544..567S} detected downflows with supersonic velocities in the \he{} lines in addition to a component nearly at rest, in a plage region close to a pore observed with the German Vacuum Tower Telescope. The inflowing material exhibited a constant acceleration of 200\,ms$^{-2}$. They interpreted this constant acceleration as a free-fall of the matter either unobstructed by magnetic structures or along vertical field lines, from the height of formation of \ion{He}{I} toward the photosphere. We see a significant fraction of the supersonic downflows to be associated with plage regions. More than half of these are located in the plage regions anchored at the footpoints of \ion{He}{I} loops and filament barbs, and show a smooth increase in the flow velocity toward the footpoints while the rest appear scattered. The chromospheric structure associated with these scattered downflows could not be identified. 

The downflows in loops or AFS are usually interpreted to be due to the draining of the plasma from the emerging flux tubes \citep{1993ASPC...46..471C,2003Natur.425..692S,2007A&A...462.1147L,2010A&A...520A..77X,2016AN....337.1050B,2018A&A...617A..55G}. Gravity and collisions with ionized particles cause the neutral \he{} gas to drain along the magnetic field lines toward the surface. A pictorial representation of this phenomena is presented in Fig. 13 of \citet{2010A&A...520A..77X} and in Fig. 15 of \citet{2018A&A...617A..55G}. In dataset 8, which shows an arcade of loops in the \ion{He}{I}c line center image, the downflows are clearly seen to follow the loops. Upflows with subsonic velocities are seen at the loop tops for the slow component (see \fig{fig:invmaps}). Faster downflows are seen closer to the footpoints suggesting gravitational acceleration of the material as it falls along the loop. In one of the loops, an asymmetry in the downflow velocities is clearly visible. Flow velocities are higher in one footpoint compared to the other. Such asymmetries were reported by \citet{2020ApJ...890...82G} and they attribute it to the difference in the photospheric magnetic field strength at the two footpoints \citep[see also][]{2004A&A...425..309S,2007A&A...462.1147L}. In accordance with them, we find that the magnetic field at the level of the photosphere derived from \ion{Si}{I} maps, is stronger at the footpoint showing faster downflows compared to the other footpoint hosting slower downflows. Even in the DARs, for example in datasets 5 and 6, the loops hosting strong downflows at the footpoints show upflows in their apex and downflows at footpoints in the slow component (see third and fourth rows in \fig{fig:app-maps2}). These signatures are in concordance with the findings of \citet{2003Natur.425..692S,2007A&A...462.1147L,2018A&A...617A..55G}. Thus, we interpret the fast downflows in loops to be a consequence of magnetic flux emergence.

One of the other proposed mechanisms for the supersonic downflows is the siphon flow mechanism along the loop either due to asymmetric heating \citep[e.g.,][]{2001SoPh..198...89B} or due to magnetic field strength differences \citep[e.g.,][]{1992A&A...261L..21R} between the footpoints. In dataset 8, where both the footpoints of the AFS are clearly visible, downflows were seen at both ends, eliminating siphon flow as their cause (siphon flow would result in upflows at one footpoint and downflows at the other). As demonstrated by \citet{1989ApJ...337..977M,1990IAUS..138..263T} siphon flow in an isolated arched flux tube can undergo a smooth transition from subsonic to supersonic speed at the top of the arch. In other datasets where we see fast downflows at the footpoints of the partly observed loops, siphon flow driven at the other footpoint that is unobserved \citep[e.g. as reported in][]{2016A&A...587A..20C} could be another channel through which the flows become supersonic.

\begin{figure}[htbp!]
    \centering
    \includegraphics[width=.48\textwidth]{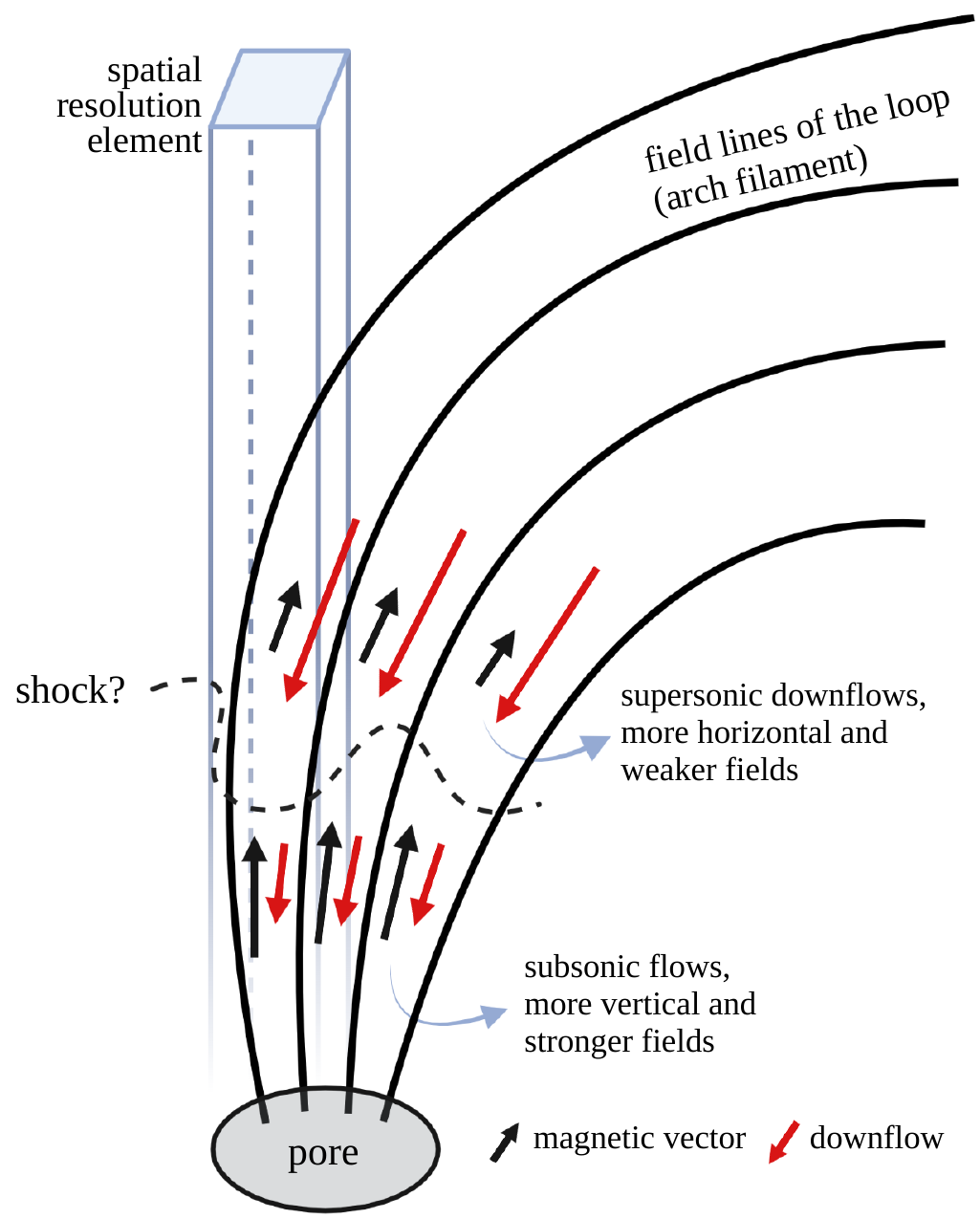}
    \caption{Proposed scenario for the most common situation in which supersonic downflows were found in the analyzed ARs, illustrated here for a positive magnetic polarity pore.}
    \label{fig:cartoon}
\end{figure}

Consequently, the favoured mechanism proposed for the supersonic downflows in He loops in AFS is that proposed by \citet{2007A&A...462.1147L}. Given the much larger statistics, we can now distinguish between their scenarios 1 and 2 (i.e. uncombed and cloud models), which they introduced to explain the coexistence of two magnetic components within a single spatial element. The fact that the field associated with the slow component in He loops is generally stronger and more vertical than the field associated with the fast component indicates that the slow component is formed below the fast component, where the field of a loop is expected to be stronger and more vertical. In between the two there is likely to be a shock, as illustrated in \fig{fig:cartoon}. Evidence for shocks has also been found in some of the datasets that we analyzed. We identified a few profiles near one of the footpoints where \ion{He}{I} exhibits weak emission, indicating heating at the footpoints. This emission could possibly be a manifestation of a shock \citep{2007A&A...462.1147L}. However, we did not optimize the inversions to fit the emission profiles and they were excluded from our analysis. Hence we cannot comment if there were supersonic flows in regions where heating was observed.
 
A \he{} filament is visible in dataset 3 \citep[this dataset has been discussed in detail in][]{2020IAUS..354..454S}. The fast downflows in this filament reach velocities of over 35\,\kms{} (see top panels in \fig{fig:app-maps2}). Mass flow in filaments along the spine and barbs have been identified in H$\alpha$ observations \citep[e.g.][]{1998Natur.396..440Z,2013SoPh..288..191J}. Redshifted flows up to 11\,\kms{} at sites where the barbs connect to the filament spine have been reported in \citet{2013SoPh..288..191J}. \citet{2013enss.confE..93P} argue that the barbs are often rooted in the intersection of about $4-5$ supergranular cells where photospheric convective motions turn into downdrafts. Following these authors, we interpret the fast downflows seen in the filament to be due to the mass motion along the filament barbs. The gravitational acceleration of the material as it falls along the barb then leads to an increase in the flow velocities away from the spine.

Multiple flows are often also found in connection with flares \citep[e.g.,][]{2003ApJ...588..596T}. There were no signs of eruptions or of any associated flares at the time the analyzed ARs were scanned, ruling out the association of these downflows with, e.g., coronal rain, i.e., material falling back toward the solar surface after a (possibly failed) eruption.

Finally, as mentioned earlier, we do find a small fraction ($\sim$8\,\%) of supersonic downflows without a coexisting slow component. This finding hints at a scenario pointed out by \citet{2000ApJ...544..567S} where the downflowing gas, present above the atmosphere where the slow component originates, pulls the slow component down during its fall. For all the remaining cases where both the slow and the fast components exist in the same resolution element, the filamentary structure of the chromosphere discussed by \citet{2007A&A...462.1147L} appears plausible.

\section{Summary and conclusions}
\label{sec:sum}
In this study, we extract some characteristics of magnetized supersonic downflows making use of the 14 spectropolarimetric observations of parts of 13 ARs in the \he{} triplet. These data are analyzed with the help of LOS velocity and magnetic field vector maps obtained by applying an inversion technique based on a multicomponent Milne-Eddington type atmospheric model. Using a set of conservative criteria, we separate the areas hosting supersonic downflows from the rest of the observed patch in each of the 14 scans. These areas are studied in detail to extract the properties of the velocity and magnetic fields in the upper chromosphere in a thin layer where the \he{} triplet forms.

Magnetized supersonic downflows are found to be anchored in all the 13 ARs, although their area coverages are only a small fraction (0.2--6.4\,\%) of the total area of the magnetized regions in the FOV. Factors such as the evolutionary phase of the AR and the quiet Sun content in the FOV of the scan influence the percentage of strong downflows that one can observe. For instance, the probability of observing strong downflows is higher during the emergence phase of an AR's evolution as emerging loops rise into the upper atmosphere and the material they are dragging up with them drains out.  Nevertheless, the fact that such flows are seen in all ARs including those in the decay phase, leads us to conclude that magnetized supersonic downflows are a common phenomena in the upper chromospheric layers of ARs.

Supersonic downflows are found in association with AR features such as pores (especially forming pores), sunspot umbrae, sunspot penumbrae, light bridges, plages, \ion{He}{I} loops (arch filaments) and filaments. The maximum value of the fast component velocity is found to lie in the range $22-49$\,\kms{}. The LOS velocities of the supersonic downflows form a broad distribution with three peaks at 14, 28, and 40\,\kms{}. Such multiple distributions are also individually exhibited by the downflows in pores, penumbrae and \ion{He}{I} loops. It is not clear if these peaks are produced by insufficient statistics or are more fundamental. In the majority of the cases (92\,\%), the supersonic downflows coexists with a second subsonic component in the same resolution element. The associated magnetic field strengths of the two components show similar distributions. However, the mean field strength is somewhat higher in the slow flow component. The inclinations suggest that the magnetic field is more vertical with respect to the solar surface for the subsonic component. Although the difference in the field strength and inclination for the two components seem to suggest that they have a different origin due to the magnetic fine structure of the upper chromosphere, in agreement with the findings of \citet{2007A&A...462.1147L}, the inclinations should be treated with caution due to the weakness of the linear polarization signals involved.

The mechanism driving these supersonic downflows appears to be manifold. A significant fraction of the supersonic downflows seem to be resulting from the draining of the material that is transported by the rising flux tubes from the solar interior to higher layers. Mass flow along barbs are responsible for the supersonic downflows in filaments. Condensation and siphon flow of the coronal mass, inverse Evershed flow, free-fall of the spicule material cold be other drivers of these strong downflows. It is possible that some of the observed supersonic downflows are chromospheric equivalents of the TR supersonic downflows. We also observed downflows for which the underlying physical cause could not be inferred. Further analysis with ARs observed in many layers of the solar atmosphere is necessary to understand this intriguing phenomenon and its relation to flows in other layers of the solar atmosphere.

\begin{acknowledgements}
We thank L. P. Chitta for useful discussions and J. de la Cruz Rodr\'iguez for providing the data from 01 June 2015. This project has received funding from the European Research Council (ERC) under the European Union’s Horizon 2020 research and innovation programme (grant agreement No. 695075) and has been supported by the BK21 plus program through the National Research Foundation (NRF) funded by the Ministry of Education of Korea. K.S. received funding from the European Union's Horizon 2020 research and innovation programme under the Marie Sk{\l}odowska-Curie grant agreement No. 797715, for a part of this project. J.S.C.D. was funded by the Deutscher Akademischer  Austauschdienst (DAAD) and the International Max Planck Research School (IMPRS) for Solar System Science at the University of G\"ottingen. The 1.5-meter GREGOR  solar telescope was built by a German consortium under the leadership of the Leibniz Institut f\"ur Sonnenphysik in Freiburg with the Leibniz Institut f\"ur Astrophysik Potsdam, the Institut f\"ur Astrophysik G\"ottingen, and the Max-Planck Institut f\"ur Sonnensystemforschung in G\"ottingen as partners, and with contributions by the Instituto de Astrof\'sica de Canarias and the Astronomical Institute of the Academy of Sciences of the Czech Republic. The GRIS instrument was developed thanks to the support by the Spanish Ministry of Economy and Competitiveness through the project AYA2010-18029 (Solar Magnetism and Astrophysical Spectropolarimetry). The data of 25 May 2015 were acquired by P. Gomory (PI of the campaign) and H. Balthasar within the SOLARNET Transnational Access and Service (TAS) program, which was supported by the European Commission’s FP7 Capacities Program under grant agreement No. 312495. GRIS data archive, SOLARNET and HiC campaigns are duly acknowledged. This study has made use of SAO/NASA Astrophysics Data System's bibliographic services.
\end{acknowledgements}

\bibliographystyle{aa}
\bibliography{downflows}

\onecolumn
\begin{appendix}
\section{Description of the ARs and location of downflows}
\label{sec:app-desc}
\begin{figure*}[ht!]
    \centering
\includegraphics[scale=0.95,trim=0.cm 10.cm 0.1cm 4.cm,clip]{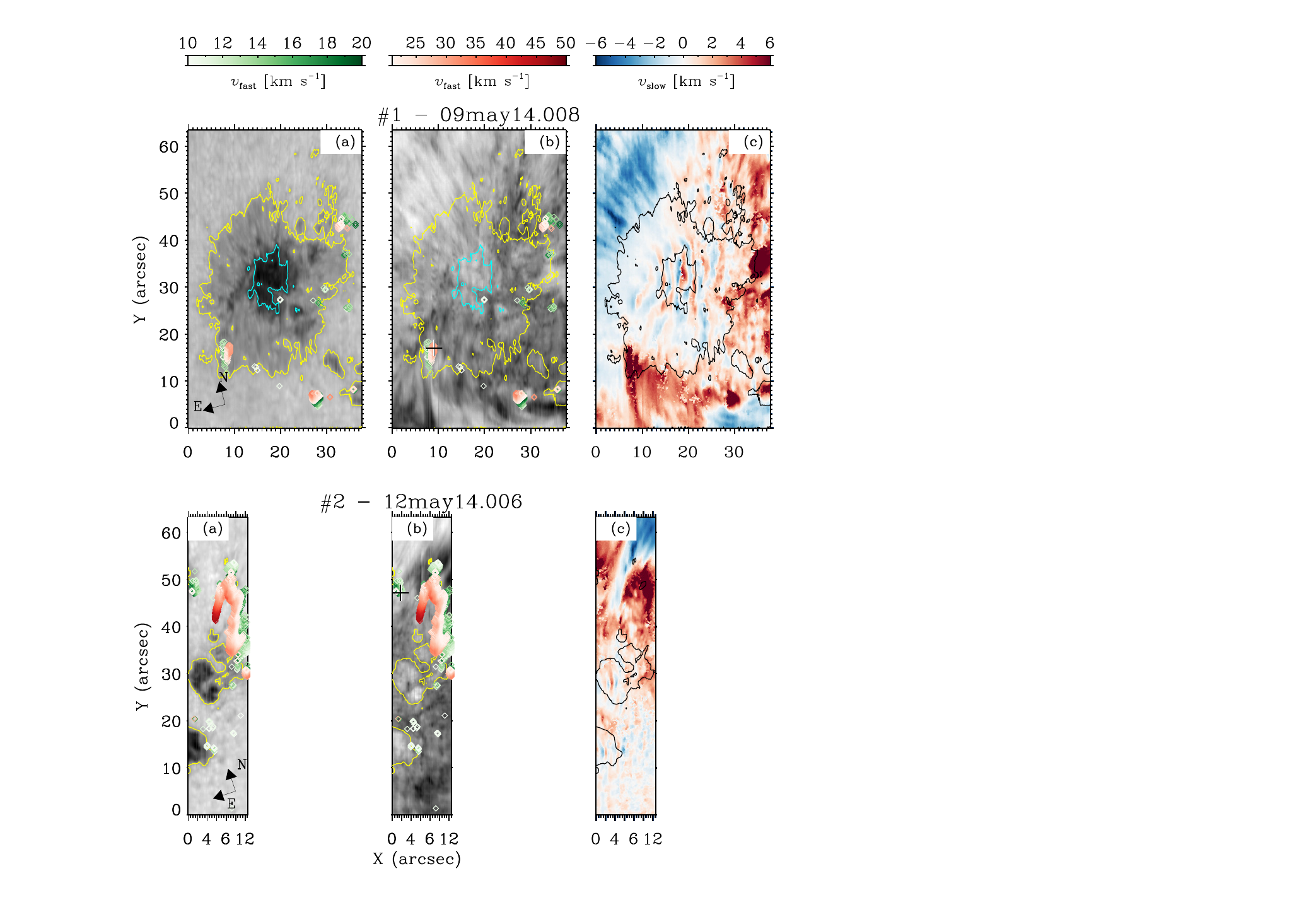}
    \caption{Overview of the ARs scanned in datasets $1-3$ as indicated in the titles (see \tab{tab:dataset}). Maps of the normalized intensities at 10825\,\AA{} (panels a) and at the \ion{He}{I}c line center (panels b), and the LOS velocity in the slow \ion{He}{I} component (panels c). The colorbars associated with panels (c) are saturated for better contrast. The arrows indicate the solar north and east directions. The yellow contours mark the boundaries of sunspots and pores, and the cyan contours mark the umbra-penumbra boundary at the photosphere. The same contours are shown in black in the LOS velocity maps of the slow component. The green and red shaded symbols overplotted on the intensity maps denote the locations of the supersonic downflows. The green symbols highlight supersonic downflows with LOS velocities up to 20\,\kms{} while red symbols represent downflows faster than 20\,\kms{} as indicated by the color bars at the top right. The black crosses in panels (b) indicate the locations of the corresponding profiles shown in \fig{fig:down}.}
    \label{fig:app-maps1}
\end{figure*}

\begin{figure*}[ht!]
    \centering
\includegraphics[scale=0.95,trim=0.0cm 2.4cm 0.1cm 0.5cm,clip]{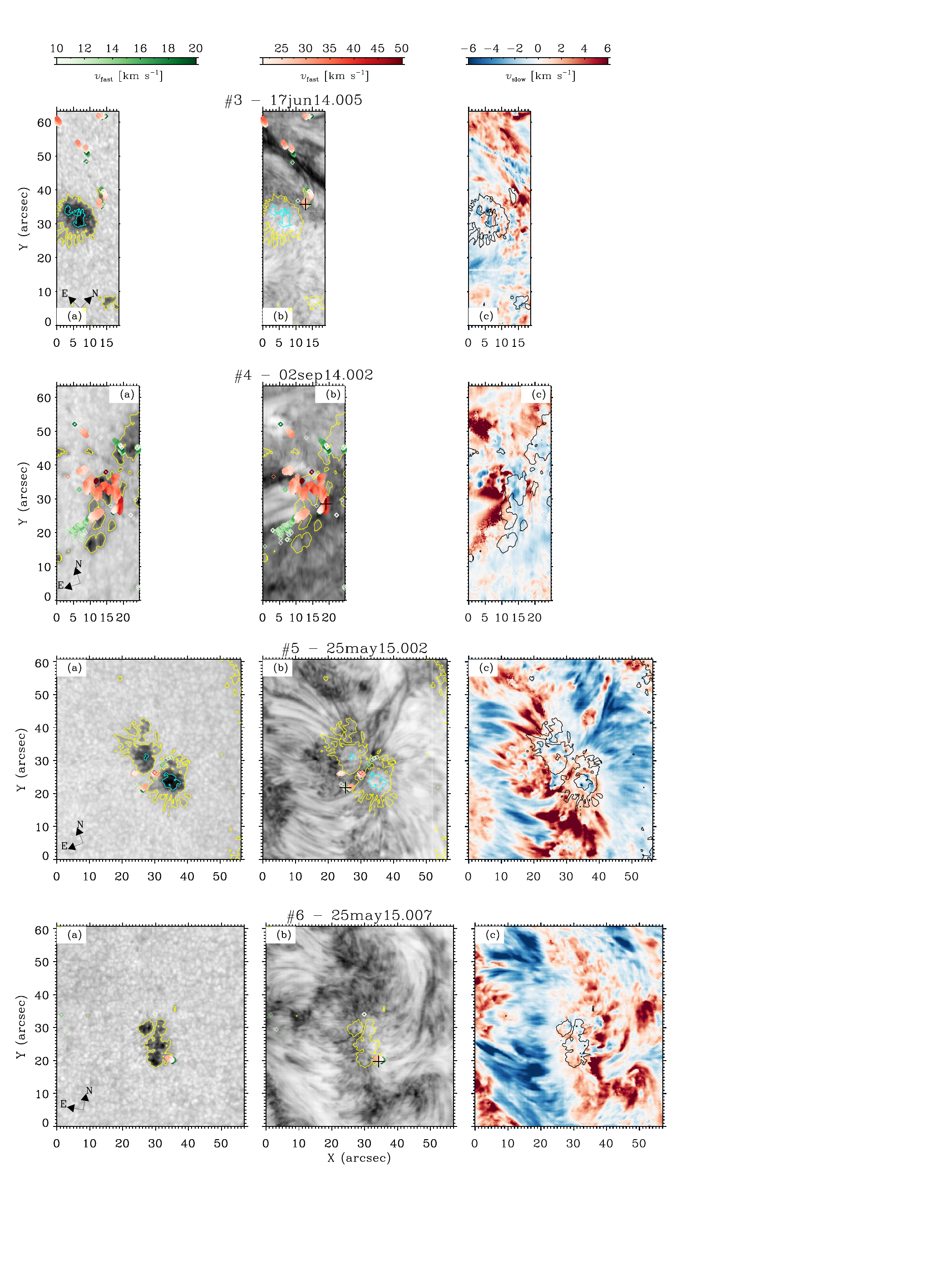}
    \caption{Same as \fig{fig:app-maps1} but for datasets $4-6$.}
    \label{fig:app-maps2}
\end{figure*}

\begin{figure*}[ht!]
    \centering
\includegraphics[scale=0.95,trim=0.0cm 2.4cm 0.1cm 1.3cm,clip]{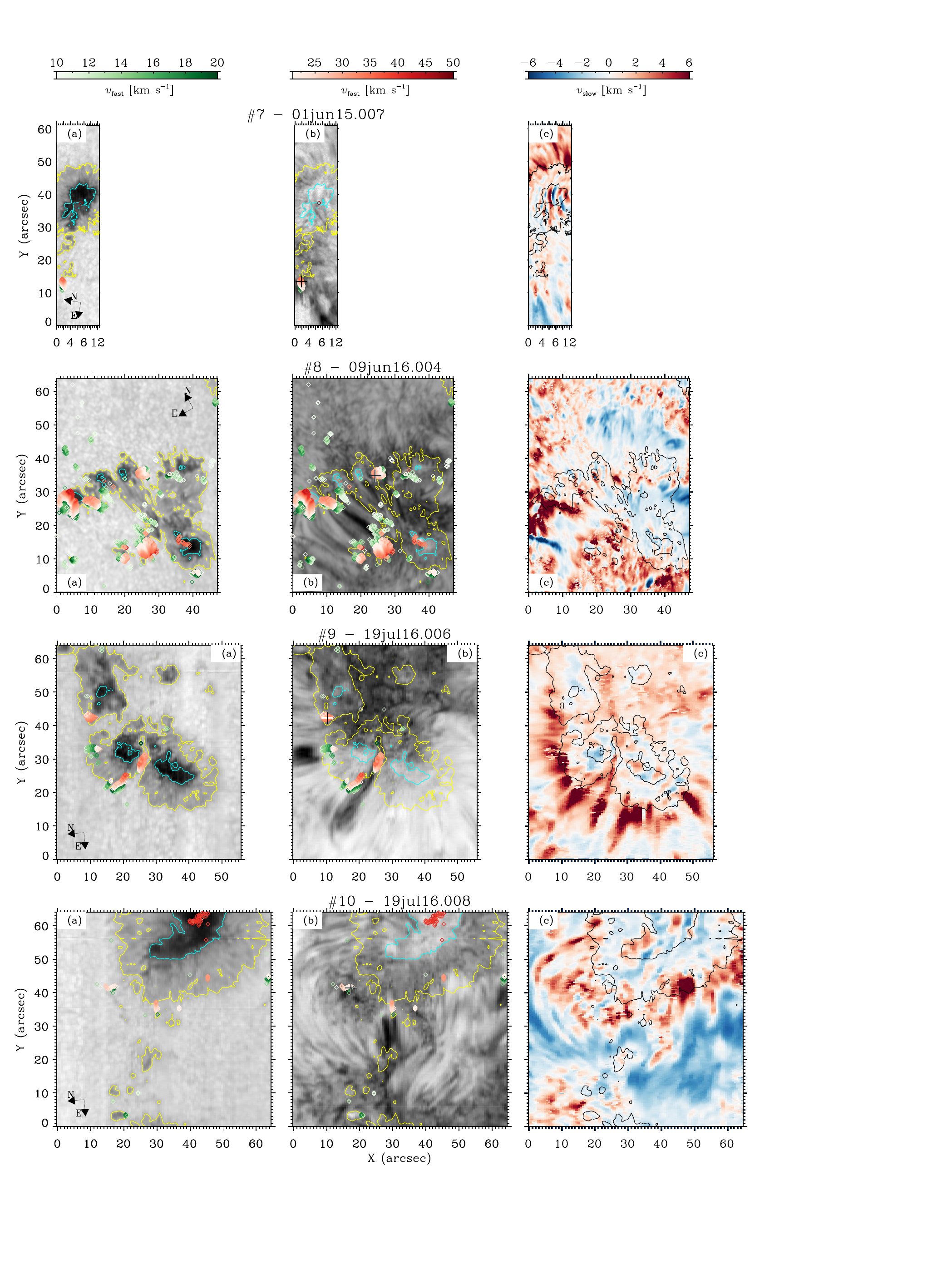}
    \caption{Same as \fig{fig:app-maps1} but for datasets $7-10$.}
    \label{fig:app-maps3}
\end{figure*}

\begin{figure*}[ht!]
    \centering
\includegraphics[scale=0.95,trim=0.0cm 3.cm 0.1cm 0.5cm,clip]{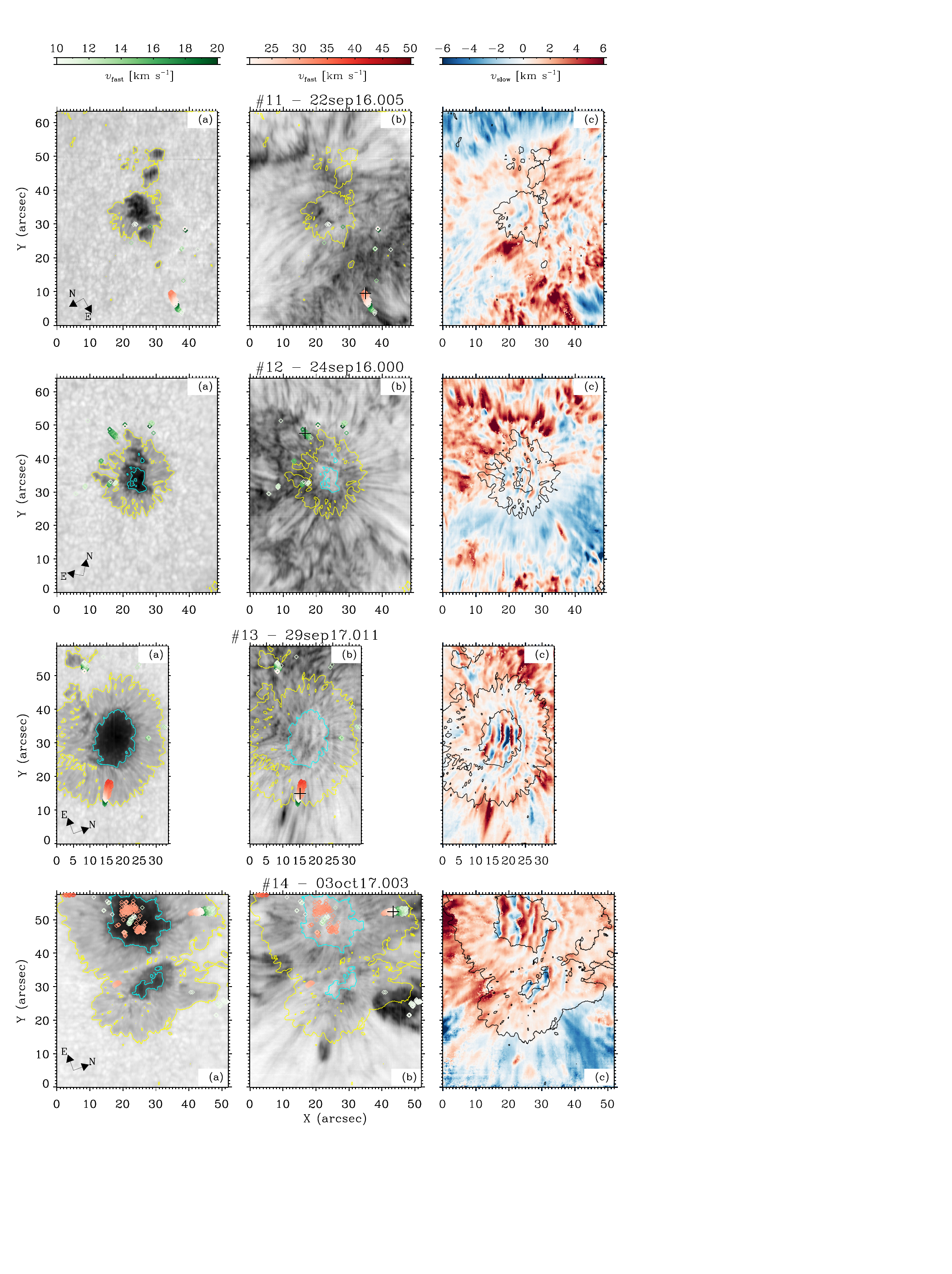}
    \caption{Same as \fig{fig:app-maps1} but for datasets $11-14$.}
    \label{fig:app-maps4}
\end{figure*}

Figs.~\ref{fig:app-maps1}--\ref{fig:app-maps4} show the raster scans of parts of all the 13 ARs in the continuum (at 10825\,\AA{}) and \ion{He}{I}c line center wavelengths, along with the LOS velocity maps for the slow \ion{He}{I} component retrieved from inversions. \he{} filaments are the dark ribbons seen in the \ion{He}{I}c line center image (dataset 3) above the regions separating the opposite polarity magnetic elements at the photosphere. In the continuum images showing sunspots with visible umbra and penumbra, the region within an intensity contour of 0.57\,$I_{\rm QS}$ (cyan contours in Figs.~\ref{fig:app-maps1}--\ref{fig:app-maps4}), where $I_{\rm QS}$ is the intensity of the surrounding quiet Sun, is considered as umbra. The region enclosed between the intensity contours of 0.57\,$I_{\rm QS}$ and 0.85\,$I_{\rm QS}$ (yellow contours in Figs.~\ref{fig:app-maps1}--\ref{fig:app-maps4}) is defined as penumbra. Note that we required an intensity threshold 0.8\,$I_{\rm QS}$ for datasets 1 and 10 and a threshold of 0.88\,$I_{\rm QS}$ for datasets 11 and 14 to define the outer penumbral boundaries (probably due to the different seeing conditions). These intensity fractions are visually derived to best represent the respective boundaries. Note that these values are slightly different from the isocontours of roughly 50\,\%{} and 90\,\%{} of quiet Sun intensity normally used to determine the inner and outer penumbra boundaries in data from space based telescopes \citep[see, for example,][]{2018A&A...611L...4J,2018A&A...619A..42L}, which is mainly due to the different wavelength (6300\,\AA{} vs. 10825\,\AA{}).



\end{appendix}

\end{document}